\documentclass[12pt,psfig,epsfig,graphicx,psfrag]{article}
\usepackage{dcolumn}
\usepackage{graphicx}
\usepackage{amssymb}
\usepackage{lscape}
\usepackage{amsmath}
\usepackage{psfrag}
\usepackage{epsfig}

\newcommand{\be}{\begin{equation}}
\newcommand{\ee}{\end{equation}}
\newcommand{\tr}{{\mathrm{tr}}}

\newcommand{\BD}{BD}
\newcommand{\AGS}{AGS}
\def\L5{\tilde{\Lambda}}

\def\Wi{W_{\infty}}
\def\Xii{\Xi_{\infty}}
\def\Z{\zeta}
\def\Wo{W_0}
\def\Xio{\Xi_0}

\def\1{\mathchoice{\rm 1\mskip-4.2mu l}{\rm 1\mskip-4.2mu l}%
{\rm 1\mskip-4.6mu l}{\rm 1\mskip-5.2mu l}}

\newcommand{\beq}{\begin{equation}}

\newcommand{\eeq}{\end{equation}}

\newcommand{\bea}{\begin{eqnarray}}
\newcommand{\eea}{\end{eqnarray}}
\newcommand{\ba}{\begin{eqnarray}}
\newcommand{\ea}{\end{eqnarray}}

\def\x{{\xi}}

\title{The Vainshtein mechanism in the Decoupling Limit of massive gravity}
\date{}
\begin{document}
\maketitle
 %\hyphenation{}

%\begin{flushright}
% %NYU-TH/02-09-11\\
%gr-qc/yymmnnn\\
%paper-20
%\end{flushright}
% \vspace{0.5cm}

\begin{center}

 %{\Large \bf  (Notes on) Spherically symmetric solutions of massive gravity in the Goldstone picture}\\
% \vspace{0.7cm}
%comment\\

 \vspace{0.3cm}
 {\large E.~Babichev\footnote{babichev@apc.univ-paris7.fr},
C.~Deffayet\footnote{deffayet@iap.fr},
R.~Ziour\footnote{ziour@apc.univ-paris7.fr}}\\
%\vspace{0.5cm}
{\it APC\;\footnote{UMR 7164 (CNRS, Universit\'e Paris 7 Denis Diderot, CEA, Observatoire de Paris)}, 10 rue Alice Domon et L\'eonie Duquet,\\
 75205 Paris Cedex 13, France.}\\
%\vspace{0.5cm}

\vspace{0.3cm}
 \bigskip

{\bf \large Abstract}
\begin{quotation}\noindent
We investigate static spherically symmetric solutions of nonlinear massive gravities. We first identify, in an ansatz appropriate to the study of those solutions, the analog of the decoupling limit (DL) that has been used in the Goldstone picture description.  We show that the system of equations left over in the DL has regular solutions featuring a Vainshtein-like recovery of solutions of General Relativity (GR). Hence, the singularities found to arise integrating the full nonlinear system of equations are not present in the DL, despite the fact those singularities are usually thought to be due to a negative energy mode also seen in this limit. Moreover, we show that the scaling conjectured by Vainshtein at small radius is only a limiting case in an infinite family of non singular solutions each showing a Vainshtein recovery of GR solutions below the Vainshtein radius but a different common scaling at small distances. This new scaling is shown to be associated with a zero mode of the nonlinearities left over in the DL. We also show that, in the DL, this scaling allows for a recovery of GR solutions even for potentials where the original Vainshtein mechanism is not working. Our results imply either that the DL misses some important features of nonlinear massive gravities or that important features of the solutions of the full nonlinear theory have been overlooked. They could also have interesting outcomes for the DGP model and related proposals.

\end{quotation}

\end{center}

\section{Introduction}
The possibility that gravity is modified at very large, i.e. cosmological, distances is motivated not only by the wish to find alternative explanations to the cosmic acceleration (or even to replace dark matter \cite{Milgrom:1983ca}), but also because it enables to separate what does and does not depend on the dynamics of gravity at cosmological scales in the standard cosmological model. Unfortunately, to obtain such modifications in a consistent way is very difficult. One simple possibility is to try to give a mass to the graviton, and theories of "massive gravity" have attracted recently some attention (see \cite{Rubakov:2008nh} for a review). Among those theories, the simplest one are what can be called nonlinear Pauli-Fierz theories, i.e. theories made by nonlinearly completing the unique consistent quadratic theory for a Lorentz invariant massive spin 2, the Pauli-Fierz theory \cite{Fierz:1939ix}. Such massive gravities share some properties with more complicated constructions, such as the Dvali-Gabadadze-Porrati (DGP in the following) model \cite{Dvali:2000hr}. The latter model has also the advantage to produce a late time acceleration of the Universe without the need for a non vanishing cosmological constant \cite{ced} (see however \cite{DEBATE1,DEBATE2,DEBATE1bis}). One of the crucial issues to be answered by such models is how to recover metrics sufficiently close to the ones obtained in Einstein's General Relativity (GR) to pass standard tests of the latter theory, while at the same time having significant deviations from GR at large distances. A first major obstacle is related to the so-called van Dam-Veltman-Zakharov (vDVZ) discontinuity  \cite{vanDam:1970vg}, i.e. the fact that the quadratic Pauli-Fierz theory does not have linearized GR as a limit when the mass of the graviton is sent to zero. Soon after the discovery of the vDVZ discontinuity, it was realized that the discontinuity might in fact disappear in nonlinear Pauli-Fierz theories \cite{Vainshtein:1972sx}. This because a careful examination of static spherically symmetric solutions of those theories by A. Vainshtein showed that the solutions of the linearized nonlinear Pauli-Fierz theories (i.e. those of simple Pauli Fierz theory) were only valid at distances larger than a distance scale, the Vainshtein radius $R_V$, which goes to infinity when the mass of the graviton is sent to zero. On the other hand, Vainshtein showed that there exists a well behaved (as the mass of the graviton is sent to zero) expansion valid at distances smaller than $R_V$, this expansion being defined as an expansion around the Schwarzschild solution of GR. What Vainshtein did not show is the possibility to join together those two expansions as expansions of one single non singular underlying solution \cite{Boulware:1973my}. In fact, in the context of nonlinear massive gravity, such a joining seems to be problematic \cite{Damour:2002gp,Jun:1986hg} even though it might work for other models where the same problem appears, such as the DGP model \cite{Deffayet:2001uk,APPROX}. Indeed Damour {\it et al.} found in Ref. \cite{Damour:2002gp} by numerical integration of the equations of motion, that singularities always appear (at least for the kind of mass term considered there) in static spherically symmetric solutions of the kind considered by Vainshtein. A common wisdom is that those singularities could be related to the instabilities known to exist in nonlinear Pauli-Fierz theories and discovered by Boulware and Deser \cite{Boulware:1973my}. Those instabilities are also believed \cite{Creminelli:2005qk,Deffayet:2005ys} to be related to the higher derivative operators appearing in a powerful effective description of the scalar sector of massive gravity proposed by Arkani-Hamed {\it et al.}  in analogy with the St\"uckelberg "trick" of gauge theories \cite{Arkani-Hamed:2002sp} introducing "Goldstone modes" and taking a "Decoupling Limit" (DL) in the theory. This limit is aimed at focusing on the strongest self-interactions of the model which are seen to arise generically in the scalar sector \cite{Arkani-Hamed:2002sp}. It has in particular the property to keep the Vainshtein radius fixed and allows to understand easily the Vainshtein mechanism as well as the scaling of the dominant terms for spherically symmetric solutions in a range of distances between the Schwarzschild radius and the Vainshtein radius \cite{Arkani-Hamed:2002sp,Creminelli:2005qk,Deffayet:2005ys}.  Our goal in this paper is to investigate the success or failure of the Vainshtein mechanism in this DL. It will shed light on the possible links between the failure of the Vainshtein mechanism in nonlinear massive gravities (as discovered by Damour {\it et al.} \cite{Damour:2002gp}) on the one hand, the Boulware Deser instability \cite{Boulware:1973my} and the Goldstone picture (as introduced in Ref. \cite{Arkani-Hamed:2002sp}) on the other hand. Clarifying those links matters not only for a better understanding of nonlinear massive gravities (theories which are probably of limited interest as far as their application to the real world is concerned) but also for the one of more sophisticated models, such as the DGP gravity, degravitation \cite{DEGRAV} or the recently proposed Galileon \cite{Nicolis:2008in}.  
For example in DGP gravity, studies of the Vainshtein recovery of GR is mostly based on a perturbation theory approach \cite{Deffayet:2001uk,APPROX} and it matters to know if it persists if one takes into account as completely as possible the full nonlinear structure of the theory (see e.g. \cite{DEBATE1bis,NONPERT}). 

The paper is organized as follows. In the next section (section \ref{MGGP}), we introduce with some details nonlinear Pauli-Fierz theories we will be interested in, the Goldstone description of the latter, and the associated Decoupling Limit. This section, mostly introductory, will nonetheless also contain some new material concerning the Goldstone picture and the DL. We then (section \ref{SSSS}) introduce appropriate ans\"atze for describing static spherically symmetric solutions of the theories introduced in the first section, as well as describe with some details the Vainshtein mechanism. Next (section \ref{SEC DL}) we turn to study static spherically symmetric solutions in the DL. We first show how to obtain the DL with the ansatz considered, then we study solutions of the DL equations of motion. Our main results appear there. We find that 
 despite the higher derivative nature of the operator appearing in the DL, there exist non singular solutions of the vacuum equations of motion in the DL which have the right large distance behaviour. We also show that, depending on the potential, this large distance behaviour (at distances larger than the Vainshtein radius) when expressed in term of a power serie expansion, might not be enough to select a unique solution at small distance. This is in fact a blessing since it leaves more room for matching a source at small distance. We also investigate sources and show how our non singular vacuum solutions can be extended inside. In addition, we show that the small distance behaviour conjectured by Vainshtein is not the only possible one. Indeed we find a more generic scaling associated with a zero mode of the nonlinear part of the DL equations of motion which also allows the recovery of GR at small distance. Interestingly, it allows a working Vainshtein-like mechanism for potentials where it was not believed to work. Our investigations (which results are summarized in section \ref{CONCLUSIONS}) also show that the instabilities seen by Damour {\it et al.} \cite{Damour:2002gp} in the full nonlinear case (i.e. not the DL), and hence the failure of the Vainshtein mechanism, are not, in contrast to a widespread belief, related to the presence of a ghost (or higher derivatives in the DL picture) in the model.

\section{Massive gravity and the Goldstone picture}
\label{MGGP}
\subsection{Massive gravity and bigravity theories}
It is well known that the only consistent Lorentz invariant mass term for a  spin two field $h_{AB}$ over a Minkowski space-time with canonical metric $\eta_{AB}$ takes the Pauli-Fierz form given by  \cite{Fierz:1939ix}
\ba \label{PFMASS}
S_{m} &=& 
-\frac{1}{8} m^2 M_P^2  \int d^4 x  h_{AB} h_{CD} \left(\eta^{AC}\eta^{BD}- \eta^{AB}\eta^{CD}\right)
\ea
where the kinetic part of the action is obtained by expanding the Einstein-Hilbert action at quadratic order in $h_{AB}$ over the flat metric $\eta_{AB}$. In the equation above $m$ is the graviton mass and $M_P$ the reduced Planck mass, given by
\be
M_P^{-2} = 8 \pi G_N,\label{Newton}
\ee
in term of the Newton constant $G_N$.
Wishing to give a nonlinear generalization of the Pauli-Fierz theory we have just defined, it seems natural to consider theories with two metrics, $g_{\mu \nu}$ and $f_{\mu \nu}$ on a four dimensional space-time\footnote{For reasons that will appear in the following, we use both capital latin letters and greak letters to indicate space-time indices.}, where 
one of the two metrics, say $g_{\mu \nu}$, will be dynamical, while the other,  $f_{\mu \nu}$, will not\footnote{Theories of this kind have been considered in the past in the context of strong interactions \cite{BIGRAV}. Note that this is not the only possible way to define a nonlinear completion of Pauli-Fierz theory, see e.g. the recent proposal \cite{NAIR}.}. Hence, we will consider the action given by
\begin{equation}
\label{action} S=\int d^4 x\sqrt{-g}\ \left(\frac{M_P^2}{2} R_g +L_g\right)  + S_{int}[f,g],
\end{equation}
In the above action, $L_g$ denotes a generic matter Lagrangian with a minimal coupling to
the metric $g$ (and not to the metric $f$), and $S_{int}[f,g]$ is an interaction term with non derivative couplings between the two metrics.   There is much freedom in the choice of this
interaction term. For example, the following two possibilities have been considered respectively 
by Boulware and Deser (\BD) in Ref. \cite{Boulware:1973my} and  by Arkani-Hamed {\it et al.} (\AGS) in Ref. \cite{Arkani-Hamed:2002sp},
\bea
S_{int}^{(2)}&=&-\frac{1}{8} m^2 M_{P}^{2}\int d^{4}x \; \sqrt{-f}\; H_{\mu \nu} H_{\sigma \tau}\left(f^{\mu\sigma}f^{\nu\tau}-f^{\mu\nu}f^{\sigma\tau}\right) \label{S2}\\
S_{int}^{(3)}&=&-\frac{1}{8} m^2 M_{P}^{2}\int d^{4}x \; \sqrt{-g}\;H_{\mu \nu} H_{\sigma \tau}\left(g^{\mu\sigma}g^{\nu\tau}-g^{\mu\nu}g^{\sigma\tau}\right) \label{S3},
\eea
where $f^{\mu \nu}$ and $g^{\mu \nu}$ denotes respectively the inverse of the metric $f_{\mu \nu}$ and $g_{\mu \nu}$, and $H_{\mu \nu}$ is defined by
\ba
H_{\mu \nu} = g_{\mu \nu} - f_{\mu \nu}.\nonumber
\ea
Following the notations of Damour {\it et al.} \cite{Damour:2002ws}, these interaction terms are in the form
\be \label{INT}
S_{int}^{(a)} = -\frac{1}{8} m^2 M_{P}^{2} \int d^{4}x {\cal V}^{(a)} (g,f) \equiv  -\frac{1}{8} m^2 M_{P}^{2} \int d^{4}x \sqrt{-g}V^{(a)} ({\bf g^{-1} f})
\ee
with ${\cal V}^{(a)}(g,f)\equiv \sqrt{-g}\;V^{(a)}({\bf g^{-1} f})$ a suitable "potential" density.
In this work we will consider theories where the potential ${\cal V}^{(a)}$ will not be necessarily one of the two above forms (\ref{S2}-\ref{S3})\footnote{We keep the numbering of the potentials (\ref{S2}-\ref{S3}) of Ref. \cite{Damour:2002ws}.}, but will be chosen such that (i) it is a scalar density under diffeomorphisms (common to the two metrics) and (ii) that when one expands $g$ to second order around the canonical Minkowski metric $\eta_{AB}$
as $g_{AB} = \eta_{AB} + h_{AB}$ and let $f$ assumes the canonical Minkowski form $\eta_{AB}$, the potential at quadratic order for $h_{AB}$ takes the Pauli-Fierz form (\ref{PFMASS}). Note however that many choices can be made with those two same properties. The theories considered will then be invariant under common diffeomorphisms which transforms the metric as
\be
\label{Ginvg}
\begin{aligned}
g_{\mu \nu}(x) &= \partial_\mu x'^{\sigma}(x) \partial_\nu x'^{\tau}(x) g'_{\sigma \tau}\left(x'(x)\right)\;, \\
f_{\mu \nu}(x) &= \partial_\mu x'^{\sigma}(x) \partial_\nu x'^{\tau}(x) f'_{\sigma \tau}\left(x'(x)\right)\;,
\end{aligned}
\ee
and under which the quantity $V^{(a)}$ transforms as a scalar. It is then possible to show  that it only depends on the matrix ${\bf g^{-1} f}$ \cite{Damour:2002ws}, hence the notation in equation (\ref{INT}). 
The equations of motion, derived from action (\ref{action}), read
\be \label{EQMot}
M_P^2 G_{\mu \nu} =\left(T_{\mu \nu}+ T^g_{\mu \nu}\right),
\ee
where $G_{\mu\nu}$ denotes
the Einstein tensor computed with the metric $g$,
$T_{\mu \nu}$ is the energy momentum tensor of matter fields, and
$T^g_{\mu \nu}$ is the effective energy momentum tensor coming from the variation with respect to the metric $g$ of the interaction term $S_{int}$. It depends non derivatively on both metrics $f$ and $g$ and is defined as usual as
\be \label{DEFTMN}
T^{g}_{\mu \nu}(x) = - \frac{2}{\sqrt{-g}} \frac{\delta}{\delta g^{\mu \nu}(x)} S_{int}(f,g).
\ee
A simple, but non trivial, consequence of equations (\ref{EQMot}) is obtained by taking a $g$-covariant derivative $\nabla$ of both sides of the equations; one gets, using the Bianchi identities and the conservation of the matter energy momentum tensor, the constraint
\ba \label{BIAN}
\nabla^\mu T_{\mu \nu}^g =0
\ea
which the effective energy momentum tensor should obey.

\subsection{The Goldstone picture}
\label{GoPi}
 The gauge invariance (\ref{Ginvg}) can be used to write the background flat metric $f$ in various coordinate systems. Starting from  a given gauge, with coordinate $X^A$,
 and the $f$ metric in the form of $f_{AB}(X)$, it might be desirable to change the gauge, but keep the change of coordinate explicit in the $f$ metric. Namely, the action is the same as action (\ref{action}), but with $f_{\mu \nu}(x)$ now given by  the expression
\ba \label{STUCA}
f_{\mu \nu}(x) &=& \partial_\mu X^A(x) \partial_\nu X^B(x) f_{AB}\left(X(x)\right),
\ea
while $g$ is kept as $g_{\mu \nu}(x)$. The quantities $X^A$, which then appear explicitly in the action of the theory,
can be considered as a set of four new dynamical scalar fields, which are analogous to the St\"uckelberg field used to restore gauge invariance in the Proca Lagrangian.
Indeed, under a subsequent coordinate change $x \rightarrow x'(x)$, $\partial_\mu X^A(x)$ transforms as
\ba
\partial_\mu X^A(x) = \partial_\mu x'^\sigma \partial_\sigma X^A(x'),\nonumber
\ea
and hence the quantity $f_{\mu \nu}(x)$ transforms as a covariant tensor of rank 2, as it should, while $f_{AB}$ is left unchanged by this coordinate transformation.
With this in mind and the analogy with the St\"uckelberg procedure and Goldstone equivalence theorem, the initial gauge, where $g$ and $f$ assume the form $g_{AB}$ and $f_{AB}$ is usually called a "unitary gauge"\footnote{In the following, we will always use 
indices with capital latin letters from the beginning of the alphabet $A,B,C,...$ to denote quantities written in a unitary gauge, like the coordinates $X^A$ and metrics $f_{AB}$ and $g_{AB}$, and greak letters from the middle of the alphabet, $\mu, \nu, ...$ to designate quantities in a non unitary gauge.}, i.e. one where the St\"uckelberg fields $X^A$ are gauged away. Note that the metric $f_{\mu \nu}$ in a non unitary gauge  can also be thought as the pullback, via the "link field" $X^A(x)$, on the space-time manifold $m_4$, with coordinates $x^\mu$, of the metric $f_{AB}$ living in an other abstract manifold ${\cal M}_4$ with coordinates $X^A$.
Usually, the unitary gauge is chosen such that in this gauge the metric $f_{AB}$ takes the canonical Minkowski form ${\rm diag}(-1,1,1,1) \equiv \eta_{AB}$. In the non unitary gauge, the action  (\ref{action}) is one for a theory with $g_{\mu \nu}$ and $X^A$ as dynamical fields. Obviously, the equations of motion for $g^{\mu \nu}$ lead to the same equations as in (\ref{EQMot}) where $f_{\mu \nu}$ is given in the form (\ref{STUCA}). Let us briefly turn to the equations of motion for $X^A$. The link fields $X^A$ only enter the action through the metric $f_{\mu \nu}$ given by equation (\ref{STUCA}), hence the variation of the action with respect to $X^A$, $\delta S_{int}(f,g)$, is given by 
\ba
\nonumber 
\delta S_{int}(f,g)= -\frac{1}{8} m^2 M_P^2\int d^4 x  \left(\frac{\delta}{\delta f_{\mu \nu}(x)} {\cal{V}}^{(a)}(f,g)\right) \delta f_{\mu \nu},   
\ea
where $\delta f_{\mu \nu}$ is the variation of $f_{\mu \nu}$ under $X^A(x) \rightarrow X^A(x) + \delta X^A(x)$. It is easily seen that the latter is given by (see appendix \ref{appA})
\be \label{dfmunu}
\delta f_{\mu \nu} = \partial_\mu \delta x^\sigma f_{\sigma \nu} + \partial_\nu \delta x^\sigma f_{\mu \sigma} + \delta x^\sigma \partial_\sigma f_{\mu \nu},
\ee
where $\delta x^\mu(x)$ is defined by
\be \label{defdxmu}
\delta x^\mu(x) = \delta X^A(x) \partial_A x^\mu(X(x))
\ee
and $x^\mu(X)$ denote the inverse mapping of $X^A(x)$.
Hence, the variation of $f$ takes the form of the Lie derivative ${\cal L}_{\delta x} f_{\mu \nu}$ of $f$ along the quantities $\delta x^\mu$ considered as a vector field on the space-time manifold.
However, the interaction term $S_{int}(f,g)$ has been constructed to be a scalar under coordinates changes, hence in particular under those of the form $x^\mu \rightarrow x^\mu + \delta x^\mu$, with $\delta x^\mu$ defined as above.
So one has
\be \label{INTVAR}
\int d^4 x \left[ \left(\frac{\delta}{\delta f_{\mu \nu}(x)} {\cal{V}}^{(a)}(f,g)\right) {\cal L}_{\delta x} f_{\mu \nu}  + \left( \frac{\delta}{\delta g_{\mu \nu}(x)} {\cal{V}}^{(a)}(f,g)\right){\cal L}_{\delta x} g_{\mu \nu} \right]  =0.\nonumber
\ee
Using this, and the definition (\ref{DEFTMN}), we get the following expression for the variation $\delta S_{int}(f,g)$ of the interaction term $S_{int}(f,g)$
\be \label{defin}
\delta S_{int}(f,g) =  \int d^4 x \sqrt{-g} \left(\nabla_\mu T^{\mu \nu}_g\right) \left(\partial_A x^\sigma\right) g_{\sigma \nu} \;\delta X^A(x).
\ee
Hence we see that the equation of motion of $X^A$ are equivalent to the Bianchi identities (\ref{BIAN}) provided that the mapping $X^A(x)$ is invertible. Note that above, we have lifted and lowered indices with $g_{\mu \nu}$. This will be the case in all the following except for what concerns the metric $f_{\mu \nu}$, the indices of which are moved with $f$ itself. Note also that in the above derivation we have not used the fact that $f_{AB}$ is the canonical Minkowski metric $\eta_{AB}$ and hence, it also holds when it is not the case.

\subsection{Strong coupling and decoupling limit in the Goldstone picture}
\label{DECSECTION}
The authors of Ref.~\cite{Arkani-Hamed:2002sp} have further developed the above mentioned analogy between $X^A$ and St\"uckelberg fields. Indeed, following \cite{Arkani-Hamed:2002sp}, one can do a "Goldstone boson" expansion of the action 
 (\ref{action}) around a unitary gauge. Indeed, considering some background solution for $g_{\mu \nu}$ (defined as $g_{\mu \nu}^0$) and $X^A(x)$ defined (the metric $f_{AB}$ being kept fixed) as 
\ba
X_0^A(x) \equiv \delta^A_\mu x^\mu,\nonumber
\ea
Ref. \cite{Arkani-Hamed:2002sp} introduces the "pion" fields $\pi^A$ as 
\ba \label{DECPIPI}
X^A(x) = X^A_0(x) + \pi^A(x),
\ea
and further does a "scalar-vector" decomposition of the $\pi^A$ in the form
\ba \label{DECPI}
\pi^A(x) = f^{AB}\left(A_B+ \partial_B \phi\right).
\ea 
Note that this notation is a bit problematic if one is interested in the structure of the theory beyond linearized equations of motion. Indeed, first neither the $\pi^A$ defined as above, nor $\phi$ and $A^B$ are tensors on the Manifold ${\cal M}_4$ (and one does not  see the need for the metric $f^{AB}$ above). Second, given the definition (\ref{DECPIPI}), it is natural to assume that $\phi$ is a function of $x^\mu$ (not of the $X^A$) and hence one does not see why it is differentiated above with respect to $X^B$.
In fact, if one takes the definition (\ref{DECPI}) literally and write $\partial_B \phi = \partial_B x^\mu \partial_\mu \phi$, one sees that $\partial_B x^\mu$ is expressible formally as a serie of $\pi^A$ (using (\ref{DECPIPI})) and hence also of $\phi$. This will generate terms at nonlinear order which will be different from the ones one gets using the naive expression 
$\partial_B \phi = \delta^\mu_B \partial_\mu \phi$. The same will be true if one considers metrics $f_{AB}$ with non trivial dependence in the coordinates $X^A$ (so that this later problem does not arise when $f_{AB}$ is taken to be the canonical flat metric, but will e.g. if one chooses to parametrize Minkowski space-time in a non trivial way). To avoid those difficulties,  
we will write, instead of (\ref{DECPI}) (and in fact this seems to be what is done implicitly in Ref. \cite{Arkani-Hamed:2002sp} to deal with nonlinear order)
\ba \label{DECPIbis}
\pi^A(x) = \delta^A_\mu\left(A^\mu(x)+ \eta^{\mu \nu} \partial_\nu \phi\right).
\ea
If one inserts the decomposition (\ref{DECPIbis}) into action (\ref{action}), and expands around flat space-time writing 
$g_{\mu \nu} = \eta_{\mu \nu} + h_{\mu \nu}$, we obtain an action for the dynamical fields $h_{\mu \nu}(x)$, $A^{\mu}(x)$ and $\phi(x)$. Since $A^\mu(x)$ and $\phi(x)$ only enter in the metric $f_{\mu \nu}$, via expression (\ref{STUCA}), the only term in action  (\ref{action}) which depends on $A^\mu(x)$ and $\phi(x)$ is the interaction term $S_{int}[f,g]$, and one has from (\ref{STUCA})
 (where no term has been neglected in the expression below)
\be
\begin{aligned}
H_{\mu \nu} &=& h_{\mu \nu} - \partial_\mu A_\nu - \partial_\nu A_\mu - 2 \partial_\mu \partial_\nu \phi \\
&& - \partial_\mu A_\sigma \partial_\nu A^\sigma - \partial_\mu \partial_\sigma \phi \; \partial_\nu \partial^\sigma \phi \\
&& - \partial_\nu A^\sigma \partial_\mu \partial_\sigma \phi - \partial_\mu A^\sigma \partial_\nu \partial_\sigma \phi. 
\end{aligned}
\label{EXPANDH}
\ee
Inserting this expression into $S[f,g]$, and keeping the lowest order in $h_{\mu \nu}(x)$, $A^\mu(x)$ and $\phi(x)$, one obtains the following first non trivial terms (upon integration by part)
\begin{equation}
\begin{aligned}
S =\frac{M_{P}^{2}}{8}&\int d^{4}x
\Big\{ 2 h^{\mu\nu} \partial_{\mu}\partial_{\nu}h - 2 h^{\mu \nu} \partial_\nu \partial_\sigma h^\sigma_\mu + h^{\mu \nu} 
 \Box h_{\mu\nu} - h \Box h\\
&+m^{2}\big[h^{2}\!-h_{\mu\nu}h^{\mu\nu}\! -F_{\mu\nu}F^{\mu\nu}\!-4(h\partial A\!-h_{\mu\nu}\partial^{\mu}A^{\nu})\!-4( h \Box \phi-h_{\mu \nu} \partial^\mu \partial^\nu \phi)\big]\Big\}\\
&+\frac{1}{2}T_{\mu\nu}h^{\mu\nu}\nonumber
\end{aligned}
\end{equation}
where $h \equiv h_{\mu \nu} \eta^{\mu \nu}$, $\partial A \equiv \partial_\mu A^\mu$, $T_{\mu\nu}$ is the matter stress-energy tensor, and indices are moved up and down with the metric $\eta_{\mu \nu}$.
The peculiarity of the above expression is that while $A^\mu$ acquires a standard kinetic term, $\phi$ does only get one 
via a mixing with $h_{\mu \nu}$ \cite{Arkani-Hamed:2002sp}, this being entirely due to the structure of the Pauli-Fierz mass term (\ref{PFMASS}). 
A may to demix $\phi$ and $h_{\mu \nu}$ is to do the shift \cite{Arkani-Hamed:2002sp}
\ba
h_{\mu \nu} = \hat{h}_{\mu \nu} - m^2 \eta_{\mu \nu} \phi.\nonumber
\ea
The quadratic interaction action then becomes
\begin{equation}
\begin{aligned}
S_{int}=\!\frac{M_{P}^{2}m^{2}}{8} \int d^{4}x\Big\{&\!
\hat{h}^{2}-\hat{h}_{\mu\nu}\hat{h}^{\mu\nu} -F_{\mu\nu}F^{\mu\nu}-4(\hat{h}\partial A-\hat{h}_{\mu\nu}\partial^{\mu}A^{\nu})\\
&+6m^{2}\left[\phi(\Box+2m^{2})\phi- \hat{h}\phi+2\phi \partial A \right]
\Big\}\nonumber.
\end{aligned}
\end{equation}
The interactions between $\phi$, $A$ and $\hat{h}_{\mu\nu}$ can be canceled by adding an appropriate gauge fixing to the action
 (see \cite{Nibbelink:2006sz}).
 Following again \cite{Arkani-Hamed:2002sp}, we can obtain canonically normalized fields $\tilde{\phi}$, $\tilde{A}$ and $\tilde{h}_{\mu\nu}$ by defining 
 \be
 \label{PHIRESCA}
 \begin{aligned}
 \tilde{h}_{\mu \nu} &=  M_P \hat{h}_{\mu \nu}, \\
\tilde{A}^\mu &= M_P m A^\mu, \\
\tilde{\phi} &= M_P m^2 \phi. 
\end{aligned}
\ee
Doing so, and expanding the action in $\tilde{\phi}$, $\tilde{A}$ and $\tilde{h}_{\mu\nu}$ one sees using (\ref{EXPANDH}) 
that $\tilde{\phi}$ has in general cubic self interactions suppressed by the energy scale
\ba \label{DEFLAMBDA}
\Lambda = \left(m^4 M_P\right)^{1/5}\;.
\ea
When those interactions are present\footnote{An appropriate choice of the interaction term $S_{inf}[f,g]$ can remove cubic (and others) self interactions of $\tilde{\phi}$ \cite{Arkani-Hamed:2002sp}.}, they are the strongest interactions among the fields $\tilde{\phi}$, $\tilde{A}$ and $\tilde{h}_{\mu\nu}$ in the limit where $m \ll M_P$, besides quadratic, cubic and quartic non derivative interactions\footnote{We include here and in the following with those interactions, a peculiar set of cubic interactions, with a derivative coupling to $\tilde A$, corresponding to $k_1+k_4=2$, $k_2=1$ and $k_3=0$ (see below), which will play a similar role.}.  Indeed, using the 
expansion (\ref{EXPANDH}), it is easy to see that those interactions, coming from the interaction term (\ref{INT}), 
scale as 
\ba \label{SELFINTCAN}
\Lambda^{4-k_1- 2 k_2 - 3 k_3 -k_4}_{k_1,k_2,k_3,k_4}  \tilde{h}^{k_1 }\left(\partial \tilde{A}\right)^{k_2} \left(\partial \partial \tilde{\phi}\right)^{k_3} \tilde{\phi}^{k_4} 
\ea
where $k_1, k_2, k_3$ and $k_4$ are integers and the expressions $\Lambda_{k_1,k_2,k_3,k_4}$
are given by 
\ba \label{L4L5}
\Lambda_{k_1,k_2,k_3,k_4} =  \Lambda \left(\frac{M_P}{m}\right)^{\frac{4k_1 + 3 k_2 + 2 k_3 + 4 k_4 -6}{5(k_1+2 k_2+3k_3+k_4-4)}}.
\ea
Whenever $k_1+2 k_2 + 3k_3+k_4 =4$ (which can only happen for $k_1+k_4=4$, $k_2=0$, $k_3=0$ or for 
$k_1+k_4=2$, $k_2=1$ and $k_3=0$), the coefficient in front of $\tilde{h}^{k_1 }(\partial \tilde{A})^{k_2} (\partial \partial \tilde{\phi})^{k_3} \tilde{\phi}^{k_4}$ in Eq. (\ref{SELFINTCAN}) is equal to the dimensionless expressions $m^2/M_P^2$ or $m/M_{P}$ and the corresponding interactions include in particular quartic non derivative couplings between $\tilde{\phi}$ and $\tilde{h}$. 
In all other cases, $\Lambda_{k_1,k_2,k_3,k_4}$ has the dimension of an energy. It is equal to $m$ for quadratic interactions and $m^2/M_P$ for cubic non derivative interactions. Hence, besides quadratic, cubic and quartic non derivative interactions, it is easily seen that the strongest interaction among the canonically normalized fields is the cubic derivative self interaction of $\tilde{\phi}$ suppressed by the scale $\Lambda = \Lambda_{0,0,3,0}$. All the other interactions are 
suppressed by a scale larger or equal to $\Lambda_{0,0,4,0}$, with 
$\Lambda_{0,0,4,0} = M_P^{1/4}m^{3/4}$ \cite{Arkani-Hamed:2002sp,Creminelli:2005qk,Nibbelink:2006sz}. In other words the exponent in the right hand side of the above formula (\ref{L4L5}) is bounded below by $1/20$ as soon as $\{k_1,k_2,k_3,k_4\} \neq \{0,0,3,0\}$ (not considering the above mentionned non derivative quadratic, cubic and quartic interactions). Hence, for $m \ll M_P$ the scale $\Lambda_4$ is stricly larger that $\Lambda$, while the non derivative cubic and quartic interactions are much smaller than the quadratic mass terms (for small expectation values of the fields). This indicates that there is a regime where the theory considered, and hence also its solutions, is well approximated\footnote{Of course the discussion here is a bit loose, it will be made more precise when dealing with spherically symmetric solutions.} by retaining only the quadratic action and the cubic self interaction of $\tilde{\phi}$, as noted in particular in Ref. \cite{Creminelli:2005qk}. This regime can be extended to arbitrarily high energy scale\footnote{We do not consider here, as done e.g. in \cite{Arkani-Hamed:2002sp}, the issue of the quantum corrections to the theory and only discuss it from a purely classical perspective.} (or arbitrarily small distances) by choosing a sufficiently large ratio $M_P/m$, as can be seen from the relation  (\ref{L4L5}).  In fact, one can take a {\it decoupling limit}, that suppresses all the interactions but the cubic $\tilde{\phi}$ derivative self interaction. This limit is
defined as 
\be
\label{DEFDEC}
\begin{aligned}
M_{P}&\rightarrow \infty, \\
m & \rightarrow  0,  \\
\Lambda &\sim {\rm constant}, \\
T_{\mu \nu}/M_P & \sim {\rm constant}. 
\end{aligned}
\ee
In this limit, all the quantities $\Lambda_{k_1,k_2,k_3,k_4}$ (with  $\{k_1,k_2,k_3,k_4\} \neq \{0,0,3,0\}$) are sent to zero or infinity. The 
action one is left with for $\tilde{\phi}$ is of the form\footnote{While $\tilde{h}$ $\tilde{A}$ become free.}
\be
\label{action_phi}
S=\frac{1}{2}\int d^{4}x\;\left\{\frac{3}{2}\tilde{\phi}\Box\tilde{\phi}+\frac{1}{\Lambda^{5}}\left[\alpha \;(\Box\tilde{\phi})^{3}+\beta\;(\Box\tilde{\phi}\;\tilde{\phi}_{,\mu\nu}\;\tilde{\phi}^{,\mu\nu})\right]-\frac{1}{M_{P}}T\tilde{\phi}\right\}\,,
\ee
where $\alpha$ and $\beta$ are numerical coefficients that can be adjusted at will by choosing an appropriate interaction term $S_{inf}[f,g]$\footnote{Note that in general, the cubic term for $\tilde{\phi}$ is given by some linear combination of the three terms 
$(\Box\tilde{\phi})^{3}$, $\Box\tilde{\phi}\;\tilde{\phi}_{,\mu\nu}\;\tilde{\phi}^{,\mu\nu}$  and $\tilde{\phi}_{,\mu\nu}\;\tilde{\phi}^{,\mu\alpha}\tilde{\phi}_{,\alpha}^{,\nu}$, but an integration by part can always be used to reduce the number of independent terms to two, as shown in Eq. (\ref{action_phi}).}, and $T$ is the trace of the energy momentum tensor.  
For example, the BD potential (\ref{S2}) leads to $\alpha=-\beta=-1/2$, while the AGS potential (\ref{S3}) leads to the opposite case $\alpha=-\beta=1/2$.
The equation of motion deriving from this action is 
\be
3\Box\tilde{\phi}+\frac{1}{\Lambda^{5}}\left[3\alpha \;\Box\left(\Box\tilde{\phi}\right)^{2}+\beta\;\Box\left(\tilde{\phi}_{,\mu\nu}\;\tilde{\phi}^{,\mu\nu}\right)+2\beta\;\partial_{\mu}\partial_{\nu}\left(\Box\tilde{\phi}\;\tilde{\phi}^{,\mu\nu}\right)\right]=\frac{1}{M_{P}}T\; .
\label{tildephi}
\ee
This equation of motion is fourth order which signals generically ghost propagating degrees of freedom. 
In fact, one can argue \cite{Creminelli:2005qk,Deffayet:2005ys} that one can see this way a generic property of massive gravity once discovered by Boulware and Deser \cite{Boulware:1973my}: namely the fact that nonlinear massive gravity propagates at nonlinear level one more degree of freedom than linear Pauli-Fierz theory, the energy of the extra propagating mode being unbounded below. This is a major obstacle to the possibility to consider nonlinear Pauli-Fierz theories defined as above as a realistic theories (see however \cite{Damour:2002wu}), but, again, our aim is here to use this theory like a toy model to study the Vainshtein mechanism, and not to advocate for a realistic use of it.

\section{Static Spherically symmetric solutions}
\label{SSSS}
In this section we introduce our framework to look for static spherically symmetric solutions of massive gravity. We also describe the Vainshtein mechanism. First, in Sec.~\ref{SEC Ansatze}, we present ans{\"a}tze for the metrics and 
discuss possible coordinate choices. 
In what follows we  will mainly be interested in the bi-diagonal ans{\"a}tze, i.e. those where both 
metrics can be put simultaneously in a diagonal form. It turns out that, in this case, 
a convenient coordinate choice is so-called the $\lambda,\mu,\nu$ gauge (following the terminology of reference \cite{Damour:2002gp}) that will be introduced. The equations of motion in this gauge (together with the Bianchi identity) form a system of ordinary differential equations for the three functions $\lambda$,
$\nu$ and $\mu$ to be determined.
This system is strongly nonlinear and cannot 
be integrated analytically. However, different asymptotic regimes can be studied separately.
First, one expansion can be made in the Newton constant, $G_N$.
This expansion will be found to be valid far from the source (i.e. for distances $R\gg R_V$, where $R_V$ is the Vainshtein 
radius that we will introduce). At lowest order, it does not match the similar expansion that one can make in General Relativity, this being reexpressed in the form of the vDVZ discontinuity. 
Vainshtein conjectured
that close to the source the General Relativity solution can be restored via the effect of the nonlinear corrections. This is the essence of the Vainshtein mechanism explained in section~\ref{SEC Vainshtein}. In the Vainshtein original proposal \cite{Vainshtein:1972sx}, a scaling with distance is proposed for the first correction to the Schwarzschild solution close to the source (i.e. for $R\ll R_V$). Here we show that there is another possible scaling that will play an important r\^ole, as will be discussed in section~\ref{SEC DL}.

\subsection{Ans\"atze and coordinate choices}
\label{SEC Ansatze}
The most general static spherically symmetric ansatz for solutions of the theory defined by action (\ref{action}) takes the form 
\ba
g_{\mu \nu} dx^\mu dx^\nu &=& - J(r) dt^2 + K(r) dr^2 +L(r) r^2 d\Omega^2,  \nonumber \\
f_{\mu \nu} dx^\mu dx^\nu &=& -C(r) dt^2 + 2 D(r) dt dr + A(r) dr^2 + B(r)  d\Omega^2\nonumber,
\ea
where $d\Omega^2 $ is the canonical metric of a unit 2-sphere
\ba \nonumber 
d\Omega^2 = d\theta^2 + \sin^2\theta d\varphi^2.
\ea
The above ansatz can further be simplified by setting $L$ be one, by a suitable choice of the radial coordinate $r$.
However, in general, it is not possible to put both metric at the same time in a diagonal form by a coordinate change without imposing restrictions on the solutions, this is because we only have one diffeomorphism invariance (\ref{Ginvg}) at hand. 
Different cases have been considered in the literature and some solutions are explicitly known in the case when metrics are not both diagonal \cite{NONDIAG,Damour:2002gp}. In this work we will only consider bi-diagonal cases where the non dynamical metric is parametrizing a Minkowski (non dynamical) background  space-time. It will turn out useful to use different type of gauge and ans\"atze.
A first gauge choice (called the "a,b,c gauge" in \cite{Damour:2002gp}) is defined by metrics in the following form
\be
\begin{aligned}
g_{AB} dx^A dx^B &= - J(r) dt^2 + K(r) dr^2 +L(r) r^2 d\Omega^2  \\
f_{AB} dx^A dx^B &= -dt^2 + d r^2 + r^2 d\Omega^2 
\end{aligned}
\label{G1} 
\ee
A simple coordinate change $X^A(X')$
\begin{align}
 z&= r\; \cos \theta, \nonumber \\
 x&= r \;\sin \theta \; \cos \varphi,  \nonumber \\
 y&= r\;\sin \theta \; \sin \varphi,\nonumber
\end{align}
puts of course $f_{AB}$ in the canonical Minkowski form $\eta_{AB}$ (we will not need the corresponding expression for $g$)
\be 
\label{G2}
f_{AB} dx^A dx^B = -dt^2 + dx^2 +dy^2 + dz ^2.
\ee
Note that both gauges (\ref{G1}) and (\ref{G2}) have in common that all the unknown functions ($J$, $K$ and $L$) are put in the metric $g$  and hence, in agreement with our definitions of section \ref{GoPi}, we can consider those gauges as unitary. 
A non unitary gauge, which has some advantages, is the one where one of the unknown function $J,K,L$ is put into the expression of the non dynamical metric, given by 
\be
\begin{aligned} 
g_{\mu \nu}dx^\mu dx^\nu &= -e^{\nu(R)} dt^2 + e^{\lambda(R)} dR^2 + R^2 d\Omega^2  \; ,\\
f_{\mu \nu}dx^\mu dx^\nu &= -dt^2 + \left(1-\frac{R \mu '(R)}{2}\right)^2 e^{-\mu(R)} dR^2 + e^{-\mu(R)}R^2 d\Omega^2\; ,
\end{aligned}
\label{lammunu}
\ee
where here and in the following a prime denotes a derivation with respect to $R$. The relation between gauges (\ref{G1}) and (\ref{lammunu}) 
is given by 
\be \label{rR}
R e^{-\mu(R)/2} = r, 
\ee
while the relation between the functions appearing in the metric coefficients are 
\ba
J(r)&=& e^{\nu(R)} , \nonumber\\
K(r) &=& e^{\lambda(R)}e^{\mu(R)}\left(1- \frac{R\mu'(R)}{2}\right)^{-2}, \nonumber \\
L(r) &=& e^{\; \mu(R)}.\nonumber
\ea
Following reference \cite{Damour:2002gp}, we will call the gauge (\ref{lammunu})
the $\lambda,\mu,\nu$ gauge. 
It has the advantage that the $g$ metric can readily be compared to the usual Schwarzschild metric of standard General Relativity. The coordinate change defined by 
\ba
Z&=& R\; \cos \theta, \nonumber \\
X&=& R \;\sin \theta \; \cos \varphi,  \nonumber \\
Y&=& R\;\sin \theta \; \sin \varphi, \nonumber 
\ea
puts then the metric $g$ and $f$ in the form 
\be
\label{G3g}
\begin{aligned}
g_{\mu \nu} dx^\mu dx^\nu &= - e^{\nu(R)} dt^2 + dX^2 + dY^2 + dZ^2 \\
& + \left(\frac{e^{\lambda(R)}-1}{R^2}\right)\left(X dX + Y dY + Z dZ\right)^2 \\
f_{\mu \nu}dx^\mu dx^\nu &= -dt^2 + e^{-\mu(R)} (dX^2 + dY^2 + dZ^2) \\
&+\left(-\frac{\mu'}{R}+ \frac{\mu'^2}{4}\right)e^{-\mu(R)}  \left(X dX + Y dY + Z dX\right)^2 %\label{G3f}
\end{aligned}
\ee
with $R^2= X^2+Y^2+Z^2$. We obtain easily the relation between coordinates $\{X^A\}=\{t,x,y,z\}$ of the unitary gauge (\ref{G2}) and coordinates $\{x^\mu\}= \{t,X,Y,Z\}$ of the non unitary gauge (\ref{G3g}) as (the time coordinate $t$ being the same in the two gauges).  
\be
\label{TRANSCART}
\begin{aligned}
x&=& X e^{-\mu(R)/2} = X \frac{r}{R} = r \partial_X R, \\
y&=& Y e^{-\mu(R)/2} = Y \frac{r}{R}= r \partial_Y R, \\
z&=& Z e^{-\mu(R)/2} = Z \frac{r}{R}= r \partial_Z R.
\end{aligned} 
\ee
The Jacobian $\tilde{J}$ of the transformation (\ref{TRANSCART}), $|\partial_\mu X^A|$, is given by
\be
\tilde{J}= e^{-\frac{3}{2} \mu(R)} \left(1- \frac{R}{2} \mu'(R)\right).\label{JACOB}
\ee
It does not vanish except possibly on a sphere of radius $R$ where 
\be \label{condJ}
\frac{R}{2} \mu'(R) = 1.
\ee
Obviously, a generic function $\mu$ would not define via Eq. (\ref{TRANSCART}) a one to one mapping 
over the whole Minkowski space. To obtain such a mapping, a sufficient condition is that $r(R)$ is a strictly monotonous function of $R$ which maps the real positive half line to itself with an everywhere non vanishing Jacobian. 
In this paper, because we will deal with the Decoupling Limit, we will not be careful about what is happening at the origin $r=0$. However, a necessary condition on $\mu$ is that $r(R)$ maps the origin to itself, and hence $\mu$ should verify
\be \label{lim0}
\lim_{R \rightarrow 0} R e^{-\mu(R)/2} = 0.
\ee 
Note that the monotonicity of $r(R)$ is automatic, provided that (\ref{condJ}) never holds. Given a particular solution for $\mu$, $\lambda$, and $\nu$, the above conditions could prevent the solution to be interpreted as a correct solution of the equation of motion of the theory (\ref{action}).

\subsection{The Vainshtein mechanism}
\label{SEC Vainshtein}
\subsubsection{A short introduction to the vDVZ discontinuity and the Vainshtein mechanism}
The (quadratic) Pauli-Fierz theory, with a mass term given as in Eq. (\ref{PFMASS}), is known to suffer from the van Dam-Veltman-Zakharov (vDVZ) discontinuity, i.e. the fact that when one lets the mass $m$ of the graviton vanish, one does not recover predictions of General Relativity. E.g., if one adjusts the parameters (namely the Planck scale) such that the Newton constant agrees with the one measured by some type of Cavendish experiment, then the light bending as predicted by Pauli-Fierz theory (and for a vanishingly small graviton mass) will be $3/4$ of the one obtained by linearizing GR \cite{vanDam:1970vg}\footnote{The fact it is smaller is easy to understand: the essential difference between Pauli-Fierz theory and linearized GR comes from an extra propagating scalar mode present in the massive theory. This mode exerts an extra attraction in the massive case compared to the massless case. Hence, if one wants measurements of the force exerted between non relativistic masses to agree, the coupling constant of the massive theory should be smaller than that of the massless theory. But light bending is blind to the scalar sector - because the light energy momentum tensor is traceless. Hence, provided the two theories agree on the force between non relativistic probes, the massive theory would predict a smaller light bending than the massless one.}.  One way to see this is to consider solutions of equations of motion (\ref{EQMot}) which are static and spherically symmetric and which would describe the metric around a spherically symmetric body such as the Sun. To do so, using the ansatz (\ref{lammunu}) is especially convenient because in this form the $g$ metric can be easily compared with the standard Schwarzschild solution. If one  tries to find a solution expanding in the Newton constant, as we recall in subsection \ref{EXPNEW}, one finds immediately the vDVZ discontinuity appearing in the form of a different ($m$ independent) absolute value of the coefficients in front of the first non trivial correction to flat space-time in $g_{tt}$ and $g_{RR}$ components  (neglecting the Yukawa decay by assuming the Compton wavelength of the graviton is much larger than other distances of interest). However, as first noticed by Vainshtein \cite{Vainshtein:1972sx}, the computation of the next order correction shows that the first order approximation ceases to be valid at distances to the source smaller than a composite scale, the Vainshtein radius defined by 
\ba \label{DEFVAIN}
R_V=\left(m^{-4}R_{S}\right)^{1/5},
\ea
where $R_S$ is the Schwarzschild radius of the source. 
This Vainshtein radius obviously diverges when one lets $m$ go to zero and in fact is much larger than the solar system size for a massive graviton with a Compton wavelength of the order of the Hubble radius. Hence, one can not conclude that massive gravity is ruled out based on solar system observations and results of the original works on the vDVZ discontinuity \cite{vanDam:1970vg}.  Vainshtein also showed that an expansion {\it defined}  around the standard Schwarzschild solution can be obtained (as recalled in subsection \ref{SMALLR}) that is well behaved when the mass of the graviton is sent to zero, opening the possibility of a recovery of GR solution at small distances $R$ of the source. Indeed, the domain of validity of this second expansion was shown to be $R \ll R_V$. Moreover, the correction found by Vainshtein to the Schwarzschild solution are non analytic in the Newton constant which could have explained the failure of the attempt to obtain a solution expanding in the Newton constant. This is the aim of this section \ref{SEC Vainshtein} to give more details on the so-called Vainshtein mechanism, i.e. the possibility of a non-perturbative recovery of solutions of GR, which is the bulk of the studies of this work.

\subsubsection{Equations of motion in the $\lambda, \mu, \nu$ gauge}
Let us first consider the gauge (\ref{lammunu}). In this gauge, the equations of motion (\ref{EQMot}) read
\ba
e^{\nu-\lambda}\left(\frac{\lambda'}{R} + \frac{1}{R^2}(e^\lambda-1)\right) &=& 8 \pi G_N\left(T^g_{tt}+\rho e^\nu\right), \nonumber\\ 
\frac{\nu'}{R}+\frac{1}{R^2}\left(1-e^\lambda\right) &=& 8 \pi G_N\left(T^g_{RR}+Pe^\lambda\right),\label{EQMOMO}
\ea
where the source energy momentum tensor $T_{\mu}^{\;\nu}$ is assumed to have the perfect fluid form 
\be
T_{\mu}^{\;\nu}=\text{diag} (-\rho, P,P,P),\nonumber
\ee 
with total mass 
\be
M\equiv\int_{0}^{R_{\odot}}4\pi R^{2}\;\rho\;dR.\nonumber
\ee
The matter conservation equation reads 
\be\label{MATTCONS}
P'=-\frac{\nu'}{2}(\rho+P), 
\ee
while the Bianchi identities have the only non trivial component
\ba\label{BIANCHI}
-\frac{1}{m^{2}M_{P}^{2}}\;\frac{1}{R}\nabla^\mu T_{\mu R}^g = 0.
\ea
Expanding equations (\ref{EQMOMO}) and (\ref{BIANCHI}) in power of $\lambda$, $\nu $ and $\mu$, we find 
\ba
\frac{\lambda'}{R}+\frac{\lambda}{R^{2}} +G_{tt}^{(Q)}(\nu,\lambda)+G_{tt}^{(C^>)}(\nu,\lambda)
&=&-\frac{m^{2}}{2}(\lambda+3\mu+R\mu')+  8 \pi G_N \rho \nonumber \\
&& + m^{2}Q_{tt}(\nu,\lambda,\mu) + \nu 8 \pi G_N \rho \nonumber \\
&& + m^{2}C^>_{tt}(\nu,\lambda,\mu)   \nonumber \\&&
+  \nu 8 \pi G_N \rho (e^\mu-1-\nu), 
\label{Gtt}\\
\frac{\nu'}{R}-\frac{\lambda}{R^{2}}+G_{RR}^{(Q)}(\nu,\lambda)+G_{RR}^{(C^>)}(\nu,\lambda)
&=&\frac{m^{2}}{2}(\nu+2\mu)+  8 \pi G_N P\nonumber \\
&& + m^{2}Q_{RR}(\nu,\lambda,\mu) + \lambda 8 \pi G_N P \nonumber \\
&& + m^{2}C^>_{RR}(\nu,\lambda,\mu)   \nonumber \\&&
+  \nu 8 \pi G_N P (e^\lambda-1-\lambda), \label{Grr}\\ \label{Bianchi}
\frac{\lambda}{R^{2}}-\frac{\nu'}{2R}-Q_{b}(\nu,\lambda,\mu)-C^>_{b}(\nu,\lambda,\mu)&=&0,
\ea
where $G_{tt}^{(Q)}$, $G_{RR}^{(Q)}$ and $Q_{b}$ represent respectively the quadratic (in power of $\lambda, \mu$ and $\nu$) part of $G_{tt}$, $G_{RR}$ and the Bianchi identity; $G_{tt}^{(C^>)}$, $G_{RR}^{(C^>)}$ and $C^>_{b}$ represent respectively the cubic and higher part of $G_{tt}$, $G_{RR}$ and the Bianchi identity. The expressions of some of those quantities can be found in appendix \ref{appB}.

\subsubsection{Expansion in the Newton constant}
\label{EXPNEW}
We can first look for solution of the system (\ref{Gtt}-\ref{Bianchi}) expanding the solution into powers of the usual Schwarzschild radius $R_S$ of the source (or into the Newton constant). More explicitly, we expand as in 
\bea
\lambda&=&\lambda_{0}+\lambda_{1}+...\nonumber \\
\nu&=&\nu_{0}+\nu_{1}+...\nonumber \\
\mu&=&\mu_{0}+\mu_{1}+... \nonumber
\eea
where $\lambda_{i},\nu_{i},\mu_{i}$ are expected to be proportional to $G_N^{i+1}$, and we consider a regime where $
\lambda_{i+1 } \ll \lambda_{i}$, $\nu_{i+1 } \ll \nu_{i}$ and $\mu_{i+1 } \ll \mu_{i}$.

At linear order, far from the source, the equations of motion reduce to
\be 
\label{LIN1}
\begin{aligned}
\frac{\lambda_{0}'}{R}+\frac{\lambda_{0}}{R^{2}}&=-\frac{m^{2}}{2}(\lambda_{0}+3\mu_{0}+R\mu_{0}'),\\
\frac{\nu_{0}'}{R}-\frac{\lambda_{0}}{R^{2}}&=\frac{m^{2}}{2}(\nu_{0}+2\mu_{0}),\\
\frac{\lambda_{0}}{R^{2}}&=\frac{\nu_{0}'}{2R}.
\end{aligned}
\ee
Notice that the expansion of the Bianchi identity does not contain a term linear in $\mu$, as can be seen from equation 
 (\ref{Bianchi}), this being due to the peculiar structure of the Pauli-Fierz mass term. The exact solution of the system (\ref{LIN1}) can be found in Ref. \cite{Damour:2002gp}. However, we are interested here in the limit where $Rm\ll 1$, and want to simplify the linear system accordingly. One can see from the third equation that $\lambda_0$ and $\nu_0$ are of the same order: $\lambda_0 \sim\nu_0$, while the first or the second equation indicates that $\mu_0 \sim\lambda_0/(mR)^{2}\gg \lambda_0,\nu_0$. As a consequence, the system can be simplified to
\bea
\frac{\lambda_{0}'}{R}+\frac{\lambda_{0}}{R^{2}}&=&-\frac{m^{2}}{2}(3\mu_{0}+R\mu_{0}'), \label{lin1}\nonumber\\
\frac{\nu_{0}'}{R}-\frac{\lambda_{0}}{R^{2}}&=&{m^{2}}\mu_{0}\label{lin2},\nonumber\\
\frac{\lambda_{0}}{R^{2}}&=&\frac{\nu_{0}'}{2R} \label{lin3},\nonumber 
\eea
with the scaling (valid for $m R \ll 1$)
\be \label{scalin}
\mu_0 \gg \lambda_0 \sim \nu_0.
\ee
One can extract from the above system an equation for $\mu_0$ reading 
\be\label{eq mu large r}
3\mu_{0}+R\mu_{0}'=0, 
\ee
which can easily be solved to get
\be \label{DOMLIN}
\lambda_{0}=\frac{C_{1}}{2R},\;\; \nu_{0}=-\frac{C_{1}}{R},\;\; \mu_{0}=\frac{1}{(mR)^{2}}\frac{C_{1}}{2R}\; .
\ee
where $C_{1}$ is a constant of integration which is expected to be proportional to $G_N$ and has to be fixed by matching to the source.

At second order, we solve 
\bea
\frac{\lambda_{1}'}{R}+\frac{\lambda_{1}}{R^{2}}+G_{tt}^{(Q)}(\nu_{0},\lambda_{0})&=&-\frac{m^{2}}{2}(\lambda_{1}+3\mu_{1}+R\mu_{1}')+m^{2}Q_{tt}(\nu_{0},\lambda_{0},\mu_{0}), \label{L1}\\
\frac{\nu_{1}'}{R}-\frac{\lambda_{1}}{R^{2}}+G_{rr}^{(Q)}(\nu_{0},\lambda_{0})&=&\frac{m^{2}}{2}(\nu_{1}+2\mu_{1})+m^{2}Q_{rr}(\nu_{0},\lambda_{0},\mu_{0}) ,\label{NU1}\\
\frac{\lambda_{1}}{R^{2}}&=&\frac{\nu_{1}'}{2R}+Q_{b}(\nu_{0},\lambda_{0},\mu_{0}). \label{BQLR}
\eea
Assuming that $\nu_1$ and $\lambda_1$ are of the same order (in line with $\nu_0 \sim \lambda_0$), 
and using the scaling (\ref{scalin}) we find from (\ref{BQLR}) the scaling 
\be
\lambda_{1}\sim\nu_{1}\sim\mu_{0}^{2}\sim\frac{R_{S}}{R}\times\frac{1}{(mR)^{4}}\frac{R_{S}}{R}.\nonumber
\ee
As a consequence the above system (\ref{L1}-\ref{BQLR}) reduces to (in the $ m R \ll 1$ limit) 
\bea
\frac{\lambda_{1}'}{R}+\frac{\lambda_{1}}{R^{2}}&=&-\frac{m^{2}}{2}(3\mu_{1}+R\mu_{1}') \label{L2}\\
\frac{\nu_{1}'}{R}-\frac{\lambda_{1}}{R^{2}}&=&{m^{2}}\mu_{1} \label{NU2}\\
\frac{\lambda_{1}}{R^{2}}&=&\frac{\nu_{1}'}{2R}+Q(\mu_{0}), \label{B2}
\eea
with $Q(\mu_0) \equiv Q_{b}(0,0,\mu_0)$. 
For the interaction terms of Eq.~(\ref{S2}) and (\ref{S3}), we find respectively (see the expressions given in appendix \ref{appB}) 
\bea
Q^{(2)}(\mu)&=&\frac{\mu'^{2}}{4}+\frac{\mu\mu''}{2}+\frac{2\mu\mu'}{R} \label{Q2}\\
Q^{(3)}(\mu)&=& - Q^{(2)}(\mu). \label{Q3}
\eea
In the general case one can show (\emph{cf.} appendix \ref{appB}) that $Q(\mu)$ is given by 
\be
\label{DEFQAB}
\begin{aligned}
Q^{(\alpha,\beta)}(\mu) &= -
\frac{1}{2 R}\left\{3\alpha\left(6 \mu \mu'+2R \mu'^{2}+\frac{3}{2}R \mu \mu''+\frac{1}{2}R^{2} \mu ' \mu''\right) \right. \\  
& \left.+\beta\left(10 \mu \mu '+5R \mu'^{2}+\frac{5}{2}R \mu \mu''+\frac{3}{2}R^{2} \mu' \mu''\right)\right\},
\end{aligned}
\ee
where $\alpha$ and $\beta$ are numerical constants depending on the interaction term $S_{int}[f,g]$.

We can eliminate $\lambda_1$ and $\nu_1$ from the above system (\ref{L2}-\ref{B2})
to get an equation for $\mu_1$
\ba
3 Q(\mu_0) + r Q'(\mu_0) = -\frac{3}{4} m^2 \left( 3 \mu_1 + R \mu'_1\right). \nonumber
\ea
Hence, gathering the linear and quadratic terms in the $G_N$ expansion, we find
\bea
\nu&=&- \frac{\bar{R}_{S}}{R} +  \frac{\bar{R}^2_{S}}{R^2} \frac{n_1}{(mR)^{4}} + \mathcal{O}(\bar{R}_S^3) \label{nuLR}\\
\lambda &=& \frac{1}{2} \frac{\bar{R}_{S}}{R}+ \frac{\bar{R}^2_{S}}{R^2} \frac{l_1}{(mR)^{4}}  + \mathcal{O}(\bar{R}_S^3)\label{lLR}\\
\mu&=& \frac{1}{2 (m R)^2}\frac{ \bar{R}_{S}}{R}+ \frac{\bar{R}^2_{S}}{R^2} \frac{m_1}{(mR)^{6}} + \mathcal{O}(\bar{R}_S^3) \label{muLR}
\eea
where $n_1$, $m_1$ and $l_1$ are order one dimensionless quantities that depend on the details of the quadratic term $Q(\mu_0)$, and $\bar{R}_S$ is the Schwarzschild radius defined as usual from the "Cavendish" Newton constant $\bar{G}_N$ (defined as the one which is measured in some Cavendish-like experiment, assuming the Newtonian potential is given at leading order by the first term in the right hand side of equation (\ref{nuLR})). To obtain the expressions (\ref{nuLR}-\ref{muLR}) above, we have fixed the integration constant $C_1$ such that the Newtonian force between non relativistic pointlike bodies matches the one obtained in Newtonian theory (assuming those bodies are separated by a distance much larger than the Vainshtein radius and much smaller than the graviton Compton wavelength).
 As we said above, this requires defining the Newton constant $\bar{G}_N$  as $4/3$ of $G_N$\footnote{So that we have also $\bar{R}_S = 4/3 R_S$, where $R_S$ is the Schwarzschild radius used in the rest of this paper.} as defined by equation Eq. (\ref{Newton}). We also see the vDVZ discontinuity in the fact that the coefficient in front of $\bar{R}_S/R$ in the first term on the right hand side of (\ref{lLR}) is not equal to one (as it would be in GR). Finally, we see that this expansion is only valid for $R \gg  R_{V}$ where $R_V$ is the Vainshtein radius defined as in (\ref{DEFVAIN}).

\subsubsection{Small $R$ expansion}
\label{SMALLR}
At small $R$, $R \ll R_V$, following Vainshtein's idea, one is looking for an expansion around solutions of usual General Relativity. In other words, one looks for an expansion in power of the mass of the graviton (squared) $m^2$ and expands the functions 
$\lambda, \nu,\mu$ as in 
\be
f(R)=\sum_{n=0}^{\infty}m^{2n}f_{n}(R)\; ,\nonumber
\ee
where the $f_n$ are $m$-independent coefficients.  
By definition, the lowest order expressions of $\lambda$ and $\nu$ are given by the Schwarzschild form 
\ba \label{SCHWA}
\lambda_{0}=-\nu_{0}=-\ln \left(1-\frac{R_{S}}{R}\right)=\frac{R_{S}}{R}+\frac{1}{2}\left(\frac{R_{S}}{R}\right)^{2}+...
\ea
Keeping the lowest order term in $R_S/R$, $\lambda_0$ and $\nu_0$ are simply obtained by solving the vacuum 
linearized Einstein equations
\ba
\frac{\lambda_{0}'}{R}+\frac{\lambda_{0}}{R^{2}}&=& 0\nonumber\\
\frac{\nu_{0}'}{R}-\frac{\lambda_{0}}{R^{2}}&=& 0.\nonumber
\ea
The function $\mu_{0}$ is found using the Bianchi equation (\ref{Bianchi})
\be
\frac{\lambda_{0}}{R^{2}}=\frac{\nu_{0}'}{2R}+Q(\mu_{0}),\nonumber
\ee
where only the lowest order terms have been kept. 
Inserting in the above equation the GR solution (\ref{SCHWA}), we find that one should have 
\be \label{EQQSIMP}
Q(\mu_0) = \frac{R_S}{ 2 R^3},
\ee
where the lowest order in $R_S/R$ has been kept. Assuming, following Vainshtein, that $\mu_0$ 
can be expanded as a power of $R$, we find that the first non trivial term should be of the form 
\ba 
\mu_{0}&=& M_0 \sqrt{R_{S}/R}, \nonumber \\
& \gg & \lambda_0 , \nu_0,\label{SCAbis} 
\ea
where $M_0$ is a pure number. 
Hence, there is a first potential obstruction to the success of the Vainshtein mechanism \cite{Damour:2002gp}, namely the left hand side of equation (\ref{EQQSIMP}) must be positive definite, which is only possible for particular quadratic terms $Q(\mu_0)$. E.g. the interaction term (\ref{S2}) leads via equation (\ref{EQQSIMP}) to an imaginary $M_{0}$ and hence, for this potential the scaling proposed by Vainshtein does not work (see however below). In contrast, from the interaction term (\ref{S3}) (and hence equation (\ref{Q3})), one finds the real value $M_0=\pm \sqrt{8/9}$.

The next order for $\lambda_1$ and $\mu_1$ is given by solving equations (\ref{Gtt}-\ref{Grr}). In the limit where 
\be
R_S  \ll R \ll m^{-1},\nonumber
\ee
taking into account the scaling (\ref{SCAbis}), those equations simply reduce to 
\ba
\frac{\lambda_{1}'}{R}+\frac{\lambda_{1}}{R^{2}}&=&-\frac{m^{2}}{2}(3\mu_{0}+R\mu_{0}') \label{lin1bis}\nonumber\\
\frac{\nu_{1}'}{R}-\frac{\lambda_{1}}{R^{2}}&=&{m^{2}}\mu_{0}\label{lin2bis},\nonumber
\ea
while $\mu_1$ is obtained solving the Bianchi identity (\ref{Bianchi}) which reduces, in the limit considered, to 
\be
\frac{\lambda_{1}}{R^{2}}- \frac{\nu_{1}'}{2R} = Q_{b}(0,0,\mu_{0}+\mu_{1}),\nonumber
\ee
where in the right hand side of the above equation only the term linear in $\mu_0$ and $\mu_{1}$ are kept (cross products). 
Eventually, we obtain an expansion of the form found by Vainshtein given by
\bea
\nu&=&-\frac{R_{S}}{R} + N_1 \left(m R \right)^{2} \sqrt{\frac{R_{S}}{R}} +\mathcal{O}\left(m^{4}\right)\;, \label{nuSR}\\
\lambda&=& \frac{R_{S}}{R} + L_1 \left(m R \right)^{2} \sqrt{\frac{R_{S}}{R}} +\mathcal{O}\left(m^{4}\right)\;, \label{lSR}\\
\mu&=& M_{0} \sqrt{\frac{R_{S}}{R}} + M_1 \left(m R \right)^{2} +\mathcal{O}\left(m^{4}\right) \label{muSR}\;,
\eea
where again $M_0$, $N_1$, $L_1$ and $M_1$ are order one dimensionless numbers.
This expansion makes sense only if $m^{2}\mathcal{O}\left(R^{2}\sqrt{\frac{R_{S}}{R}}\right)\left[1+\mathcal{O}\left(\frac{R_{S}}{R}\right)\right]\ll\frac{R_{S}}{R}\left[1+\mathcal{O}\left(\frac{R_{S}}{R}\right)\right]$, \textit{i.e.}, if $R\lesssim R_{V}=\left(m^{-4}R_{S}\right)^{1/5}$.

Note however that there is another possible expansion at small $R$ that has not been discussed previously in the literature and for which one also recovers the GR solution. Indeed, the operator $Q$, appearing on the left hand side of equation (\ref{EQQSIMP}), can have a zero mode in the form of a power law, $\mu_Q \propto R^p$, and one can find an expansion for $\mu$ starting with this zero mode. E.g. in the cases of interaction terms (\ref{S2}) and (\ref{S3}), where the corresponding operators $Q$ are given by equations (\ref{Q2}) and (\ref{Q3}), one finds $p = -2$, and hence in those cases,  as we will discuss in more details in the next section (together with a discussion of the most general case) it is possible to find a solution to equation (\ref{EQQSIMP}) in the form 
\be
\mu_0 = \left(m R_S \right)^{2/5}\left( A_0 \left(\frac{R_V}{R}\right)^2 -  \frac{s}{3 A_0 } \frac{R}{R_V} \log(R/R_V)\right) + {\cal O}\left(\frac{R}{R_V}\right), \label{muSRbis}
\ee
where $s=-1$ for the potential  (\ref{S2}) and $s=+1$ for the potential (\ref{S3}), and $A_0$ is a pure number. 
Such an expansion will in fact, as will be shown in the next section, turn out to be the most general one found by numerical integration of the equations of motion in vacuum.

\subsubsection{Matching small $R$ and large $R$ behaviours}
It was noticed immediately after the seminal work of Vainshtein that there was no warranty that one could match the solution found by Vainshtein in the small $R$ limit, given by equations (\ref{nuSR}-\ref{muSR}), to the one obtained in the large $R$ limit, and given by equations (\ref{nuLR}-\ref{muLR}) \cite{Boulware:1973my}. More recently, numerical integration of the full (nonlinear) system of equations (\ref{Gtt}-\ref{Bianchi}) has shown indeed \cite{Damour:2002gp} (see also \cite{Jun:1986hg}), that integrating inwards from the large $R$ behaviour  (\ref{nuLR}-\ref{muLR}) or outwards, from the small $R$ behaviour, always results in singularities appearing at finite $R$ (provided one insists upon the solution to be asymptotically flat).  In fact the exact reason for which those singularities arise has not been clarified so far in the literature and it is one of the main purpose of this work to reexamine this question in the light of the Goldstone formalism in the decoupling limit of Ref. \cite{Arkani-Hamed:2002sp}. The general understanding is that one can link those singularities, and hence the failure of the Vainshtein mechanism, to the ghost-like nature of the extra degree of freedom discovered by Boulware and Deser \cite{Boulware:1973my} to be present in the nonlinear theory, and, as recalled above, appearing in the Goldstone formalism in the form of higher derivative operators \cite{Creminelli:2005qk, Deffayet:2005ys}.

\section{The decoupling limit in spherically symmetric solutions}
\label{SEC DL}

In this section we study static spherically symmetric solutions in the decoupling limit (DL in the following).
We first show (section \ref{Sec DL limit un}) how to obtain a DL in the case of the static spherically symmetric ansatz (\ref{lammunu}). 
We then show (\ref{Sec DL limit deux}) that this DL is in fact analogous to the one introduced in section \ref{DECSECTION} and we show the scalar field $\phi$ is closely related to the metric function $\mu$ of the gauge (\ref{lammunu}). We give the explicit link between the two and show that in the decoupling limit, both fields obey the same equation of motion. 
The DL  allows 
to decouple the scalar degree of freedom, which appears due to the 
presence of the potential term, from the degrees of freedom also present in GR. In this limit, the study of solutions is greatly simplified, in particular,
instead of three equations of motion (\ref{Gtt}-\ref{Bianchi}) there will 
be only one (nonlinear) equation for the ``gauge'' function $\mu$ (or equivalently for $\phi$). This is
consistent with the fact that by extracting the DL the  
GR degrees of freedom become decoupled from the scalar degrees of freedom.

On the other hand, one can hope that the DL will contain the 
main features of the full system, since it isolates the "troublesome" 
scalar degrees of freedom. For example, the large-$R$ asymptotic solution and 
the Vainshtein solution of the full system have DL analogues.
On the other hand, as was recalled at the end of the previous section, the Vainshtein mechanism was found to fail for the theory considered here, by integrating the full nonlinear system of equations (\ref{Gtt}-\ref{Bianchi}). Thus a natural question to ask is whether one can already see this failure in the DL. 
In the section \ref{VALIDEC}, we discuss briefly the expected range of validity of the DL.

The main results of this section are obtained in section \ref{results DL}
where we solve numerically the DL equations of motion. 
After a discussion of boundary conditions at infinity (subsection \ref{421}), we first introduce a simple linear differential equation which shares some crucial properties with the one obtained in the DL (subsection \ref{422}). Then we discuss solutions outside the source and inside the source in the simplest cases of potential (\ref{S2}) and (\ref{S3}) (subsections \ref{BDAGSout} and \ref{BDAGSin}). Last, we turn to discuss the most general case (subsection (\ref{425}).

\subsection{Extracting the limit}
\label{Sec DL limit}
\subsubsection{Rescaling and limiting equations of motion}\label{Sec DL limit un}
First, let us note, as should be clear from the derivations presented in section \ref{SEC Vainshtein}, that both the expansion in the Newton constant (in the limit where  $R \ll m^{-1}$) as given by equations (\ref{nuLR}-\ref{muLR}) and the expansion in the mass of the graviton (in the limit where $R \gg R_S$) as given by equations (\ref{nuSR}-\ref{muSR}) (as well as (\ref{muSRbis})) are left unchanged if one first does the rescaling on the metric functions,
\ba
\tilde{\nu}&\equiv& M_{P} \nu, \nonumber \\ 
\tilde{\lambda}&\equiv& M_{P} \lambda, \nonumber \\
\tilde{\mu}&\equiv&  m^{2} M_{P} \mu, \quad
\label{DEFMETSCA}
\ea
and on the components of the energy-momentum tensor
\be
\begin{aligned}
\quad\tilde{\rho} &\equiv \rho / M_P,\\ 
\tilde{P} &\equiv P/ M_P ,
\end{aligned}\nonumber
\end{equation}
and then takes the DL as it is defined in (\ref{DEFDEC}). 
More precisely, plugging the rescaled functions $\tilde{\nu}, \tilde{\lambda}$ and $\tilde{\mu}$ into the equations of motion (\ref{Gtt}-\ref{Bianchi}), and assuming those functions to be bounded, one sees that in the DL (\ref{DEFDEC}), the nonlinear system of equations (\ref{Gtt}-\ref{Bianchi}) collapses to the much simpler one 
\ba
\frac{\tilde{\lambda}'}{R}+\frac{\tilde{\lambda}}{R^{2}}&=&-\frac{1}{2}(3\tilde{\mu}+R\tilde{\mu}') + \tilde{\rho}\nonumber \\
\frac{\tilde{\nu}'}{R}-\frac{\tilde{\lambda}}{R^{2}}&=&\tilde{\mu}\nonumber \\
\frac{\tilde{\lambda}}{R^{2}}&=&\frac{\tilde{\nu}'}{2R}+\frac{Q(\tilde{\mu})}{\Lambda^{5}}
\label{tlambda}
\ea
where $\Lambda$ is defined as in (\ref{DEFLAMBDA}).
So the only non linearities which remain in this limit are those corresponding to the piece of the Bianchi identity quadratic in $\mu$. Note that we have set the pressure $\tilde{P}$ to zero in the above equations (\ref{tlambda}), since this is a direct consequence of taking the decoupling limit in the matter conservation equation (\ref{MATTCONS}):
\be\label{EQP}
\tilde{P}'=-\frac{\tilde{\nu}'}{2M_{P}}(\tilde{\rho}+\tilde{P})\rightarrow 0 \quad\text{when}\quad M_{P}\rightarrow 0\; .
\ee
It is easy to extract from the above system (\ref{tlambda}) a single equation obeyed by $\tilde{\mu}$
reading 
\be \label{EQMUT}
\frac{1}{\Lambda^{5}}\left[6Q(\tilde{\mu})+2RQ(\tilde{\mu})'\right]+\frac{9}{2}\tilde{\mu}+\frac{3}{2}R\;\tilde{\mu}'= \tilde{\rho}\;.
\ee
As we will see in the next subsection, one can map this equation to the equation of motion found for $\tilde{\phi}$, equation (\ref{tildephi}), and hence there is a very clear relation between the decoupling limit as we just defined, obtained from the equations of motions in the $\lambda, \mu, \nu$ gauge, and the one discussed in the original reference \cite{Arkani-Hamed:2002sp} as summarized in section \ref{DECSECTION}.
Before turning to discuss in detail this relation, let us first note that equation (\ref{EQMUT}) can 
easily be integrated once leading to the first integral
\be  
\frac{2}{\Lambda^{5}}Q(\tilde{\mu})+\frac{3}{2}\;\tilde{\mu}=\frac{C_{2}}{R^{3}} + \frac{1}{R^3} \int_0^R d\tilde{R}\; \tilde{\rho}\left(\tilde{R}\right) \;\tilde{R}^2 \, \label{FI}
\ee
where $C_2$ is an integration constant. 
Notice that outside the source, the integral on the right hand side of equation (\ref{FI}) gives  a constant and hence this simply results in a shift of the integration constant $C_2$ to yield 
\ba
\frac{2}{\Lambda^{5}}Q(\tilde{\mu})+\frac{3}{2}\;\tilde{\mu}=\frac{C_{0}}{R^{3}}. \label{FIbis}
\ea
As usual, the integration constant $C_0$ should be fixed by matching to the source.
The two limiting regimes discussed in section \ref{SEC Vainshtein} are easily recovered from this equation. 
The lowest order term in the expansion into the Newton constant (valid at $R \gg R_V$) corresponds to keeping only the linear order in $\tilde{\mu}$ in the left hand side of (\ref{FIbis}), recovering equation (\ref{eq mu large r}). The lowest order term in the expansion in $m^2$ (valid at $R \ll R_V$) corresponds to keeping only the quadratic term $Q(\tilde{\mu})$ in the left hand side of (\ref{FIbis}), recovering the scaling  $\mu \sim \mathcal{O}\left(\sqrt{R_{S}/ R}\right)$ found in equation (\ref{muSR}). Assuming the validity of the Vainshtein recovery mechanism, one can easily fix the (shifted) integration constant $C_0$ for a point like source, by using the known form of the functions $\tilde{\nu}$ and $\tilde{\lambda}$ at lowest order in the $m^2$ expansion which is given by their expressions in standard General Relativity (at lowest order in $R_S/R$). This fixes that, in this limit (see e.g. equation (\ref{EQQSIMP})), 
\be
\label{GRLIMIT}
\frac{2}{\Lambda^{5}}R^{3}Q(\tilde{\mu})\rightarrow M_{P}R_{S},
\ee
and hence, 
\be
C_{0}= M_{P}R_{S}\; .\label{FIXCZERO}
\ee
Note that the above equations (\ref{FI}-\ref{FIbis}) take their most simple expressions when $Q$ is given as in equations (\ref{Q2}-\ref{Q3}), i.e. for the interaction terms (\ref{S2}) and (\ref{S3}), in which case e.g. equation (\ref{FIbis}) reads 
\be
\label{eom mu}
-\frac{s}{\Lambda^{5}}\left(\frac{\tilde{\mu}'^{2}}{2}+\tilde{\mu}\tilde{\mu}''+\frac{4\tilde{\mu}\tilde{\mu}'}{R}\right)+\frac{3}{2}\tilde{\mu}=\frac{C_0}{R^{3}},
\ee
where again $s=-1$ for the BD interaction term (\ref{S2}) and $s=+1$ for the AGS mass term (\ref{S3}). For future use, and for numerical integration, it turns out convenient to introduce the dimensionless quantities 
\be
\begin{aligned}
\quad
\xi &\equiv {R}/{R_{V}}\;,\\
\rho_a &\equiv4\pi \frac{R_{V}^{3}}{M}\rho\;,\\
w(\xi) &\equiv a^{-2}\;\mu\;,\\
v(\xi) &\equiv a^{-4}\;{\nu}\;,\\
u(\xi) &\equiv a^{-4}\;{\lambda}\;,
\end{aligned}\nonumber
\end{equation}
where we used the dimensionless parameter
\be
a \equiv  R_{V} m  =(R_{S} m)^{1/5}\ .\nonumber
\ee
The functions $u$ and $v$ are just the dimensionless functions $\lambda$ and $\nu$ associated with the $g_{\mu\nu}$ metric, while $w$ corresponds to $\mu$, and $\xi$ is the dimensionless distance $R$ expressed in unit of the Vainshtein radius $R_V$. 
In term of these quantities, the system (\ref{tlambda}) reads
\ba
\frac{\dot{u}}{\xi}+\frac{u}{\xi^{2}}&=&-\frac{1}{2}(3w+\xi\dot{w}) + \rho_a\nonumber\\
\frac{\dot{v}}{\xi}-\frac{u}{\xi^{2}}&=&w\nonumber\\
\frac{u}{\xi^{2}}&=&\frac{\dot{v}}{2 \xi}+Q(w),\nonumber
\ea
where a dot denotes a derivative with respect to $\xi$. 
Eliminating $u$ and $v$ from the above system, we get the analogous of Eq. (\ref{FIbis}) for the rescaled variable $w$
\be 
\label{Eqy} 
2\;Q(w) +\frac{3}{2}w=\frac{c_0}{\xi^{3}},
\ee
where 
$c_0$ is defined by
\be
c_0 = \frac{C_0}{R_V^5 \Lambda ^5}.\nonumber
\ee
Fixing the integration constant as in (\ref{FIXCZERO}), leads to $c_{0}=1$, and the equation for $w$ now reads
\be
\label{Eqy1} 
2\;Q(w) +\frac{3}{2}w=\frac{1}{\xi^{3}},
\ee
which will be assumed from here-on and until the end of this article. The general form of $Q$ can be read from equation (\ref{DEFQAB}), $R$ being replaced by $\xi$ and $\mu$ by $w$. It is given by  
\be
\label{Qab}
\begin{aligned}
Q(w)=-\frac{1}{2}\Bigg\{&
3\alpha\left(\frac{\xi}{2}\dot w\ddot w+ 
\frac{3}{2}w\ddot w+2\dot w^2+\frac{6w\dot w}{\xi}\right)\\
&+
\beta\left(\frac{3\xi}{2}\dot w\ddot w+ 
\frac{5}{2}w\ddot w+5\dot w^2+\frac{10w\dot w}{\xi}\right)\Bigg\}.
\end{aligned}
\ee
E.g., in the simplest case of the potentials leading to equation (\ref{eom mu}), one has $\alpha+\beta=0$ and equation (\ref{Eqy1}) reads  
\be \label{Eqysimp} 
 - s  \left(\frac{\dot{w}^{2}}{2}+ w \ddot{w} + 4 \frac{w \dot{w}}{\xi}\right)+\frac{3}{2}w=\frac{1}{\xi^{3}}.
\ee
We now turn to compare the above obtained equations for $\tilde{\mu}$ (or $w$), Eq.~(\ref{Eqy}), with the equation of motion of $\tilde{\phi}$. 

\subsubsection{Comparison with the Goldstone picture}\label{Sec DL limit deux}
The most general equation of motion for $\tilde{\phi}$, Eq.~(\ref{tildephi}), can be rewritten in the
following form involving a total derivative
\be
\nabla_\mu \left\{3 \Lambda^5 \nabla^\mu\tilde\phi+
3\alpha\nabla^\mu\left(\Box\tilde\phi\right)^2+
\beta\nabla^\mu\left(\tilde\phi_{;\delta\gamma}\right)^2+
2\beta\nabla^\nu\left(\Box\tilde\phi\tilde\phi^{;\mu}_{\;\;\nu}\right)\right\}
=\frac{\Lambda^5}{M_{P}}T .\nonumber\label{tildephi1}
\ee
Hence, it is easy to see that, in the case of a spherically symmetric configuration 
$\tilde{\phi}(R)$, this equation  can always be integrated once to yield 
\ba\label{eom phi}
3\;\frac{\tilde{\phi}'}{R}&+& \frac{2}{\Lambda^{5}}\left\{3\alpha\ \left(-4 \frac{\tilde{\phi}'^{2}}{R^4} + 2 \frac{\tilde{\phi}'\tilde{\phi}''}{R^3} + 2 \frac{\tilde{\phi}''^2}{R^2} + 2 \frac{\tilde{\phi'} \tilde{\phi}^{(3)}}{R^2} + \frac{\tilde{\phi}'' \tilde{\phi}^{(3)}}{R}\right)\right. + \nonumber \\
 &&\quad\left.+\beta\left(-6\frac{\tilde{\phi}'^{2}}{R^{4}}+2\frac{\tilde{\phi}'\tilde{\phi}''}{R^{3}}+4\frac{\tilde{\phi}''^{2}}{R^{2}}+2 \frac{\tilde{\phi}' \tilde{\phi}^{(3)}}{R^2} + 3 \frac{\tilde{\phi}'' \tilde{\phi}^{(3)}}{R}
\right)\right\}\\&&\qquad\qquad\qquad\qquad\qquad\qquad\qquad\qquad=
-\frac{\tilde{C}_{2}}{R^{3}} -\frac{1}{R^3} \int_0^R d\tilde{R}\; \tilde{\rho}\left(\tilde{R}\right) \;\tilde{R}^2\nonumber,
\ea
where we used that, according to Eq.~(\ref{EQP}), $T/M_{P}=-\tilde{\rho}$ in the decoupling limit, and where $\tilde{C}_{2}$ is an integration constant. Let us now see how this equation compares with equation (\ref{FI}) obtained for $\tilde{\mu}$ in the previous subsection.  
As we have seen, the relation between the unitary gauge (\ref{G2}) and the gauge (\ref{G3g}) is encoded into the function $\mu(R)$ though the coordinate change (\ref{TRANSCART}). On the other hand, it is also contained in the Goldstone field $\pi^A$, and in our case, in a single scalar field $\phi$ defined as in (\ref{DECPIPI}-\ref{DECPI}) by
\be \label{DEFPHIMU}
\eta^{AB} \partial_B \phi = X^A(x^\mu)- \delta^A_\mu x^\mu,
\ee
where as follows from (\ref{TRANSCART}) $X^A$ and $x^\mu$ are given by $\{X^A\}=\{t,x,y,z\}$ and $\{x^\mu\}= \{t,X,Y,Z\}$, and $X^A(x^\mu)$ is given by the coordinate change (\ref{TRANSCART}). It is easy to see that equation (\ref{DEFPHIMU}) has the solution $\phi(R)$ 
\be \label{PHIMU}
\phi'\equiv\partial_{R}\phi= R \left(e^{-\frac{\mu(R)}{2}}-1\right), 
\ee
and that in this particular case no vector field $A^B$ is needed. 
Note that the relation between $\mu$ and $\phi$ is in fact "non perturbative" because of the presence of the exponential in the above relation.
The limiting conditions (\ref{lim0}) translate via relation (\ref{PHIMU}) into
\ba \label{BOUPHI}
\lim_{R\rightarrow 0} \phi'(R) &=& 0. 
\ea
This is precisely the limiting condition one would impose on a scalar field $\phi$ in a spherically symmetric configuration in order to have everywhere well defined second derivatives.   
For small $\mu$, the above expression (\ref{PHIMU}) can be expanded as 
\be
\phi'\equiv\partial_{R}\phi=-R\left(1-e^{-\frac{\mu(R)}{2}}\right)\sim -\frac{R\mu}{2}+\frac{R\mu^{2}}{8} + ...\nonumber
\ee
If we go to the canonically normalized $\tilde{\phi}$, defined in equation (\ref{PHIRESCA}), and use the rescaled metric variables as defined in (\ref{DEFMETSCA}), and then take the decoupling limit (\ref{DEFDEC}), one is left with the simple relation between  $\tilde{\phi}$ and $\tilde{\mu}$ given by 
\be \label{MUTPHIT}
\tilde{\mu}=-\frac{2}{R}\tilde{\phi}'\; .
\ee
Substituting $\tilde{\phi}'$ in the equation (\ref{eom phi}) we get then
\be
\begin{aligned}
\frac{3}{2}\tilde{\mu}-\frac{1}{R  \Lambda^{5}}&\Bigg\{3\alpha\left(6\tilde{\mu}\tilde{\mu}'+2R\tilde{\mu}'^{2}+\frac{3}{2}R\tilde{\mu}\tilde{\mu}''+\frac{1}{2}R^{2}\tilde{\mu}'\tilde{\mu}''\right)\\
&+\beta\left(10\tilde{\mu}\tilde{\mu}'+5R\tilde{\mu}'^{2}+\frac{5}{2}R\tilde{\mu}\tilde{\mu}''+\frac{3}{2}R^{2}\tilde{\mu}'\tilde{\mu}''\right)\Bigg\}=\frac{\tilde{C}_{2}}{R^{3}} +\frac{1}{R^3} \int_0^R \! d\tilde{R}\;\tilde{R}^2 \tilde{\rho}\left(\tilde{R}\right),\label{Qmutilde}\nonumber
\end{aligned}
\ee
and we recover exactly equation (\ref{FI}) (identifying $\tilde{C}_2$ and $C_2$). 
It is however interesting to note that the boundary condition (\ref{BOUPHI}) can be lost in the decoupling limit due to the "non perturbative" (exact) relation between $\phi$ and $\mu$ given by equation (\ref{PHIMU}).
For example, if one considers the field theory defined by action (\ref{action_phi}) and looks for a spherically symmetric solution, one is led to impose that $\tilde{\phi}^\prime(0)$ vanishes. This, via relation (\ref{MUTPHIT}) imposes however a condition on $\tilde{\mu}$ which is not necessary considering the conditions given at the end of section \ref{SEC Ansatze}.

\subsubsection{Expected range of applicability of the Decoupling Limit}\label{VALIDEC}
We would like here to discuss to what extent the DL is capturing the leading behaviour of the solution of the full nonlinear system. As we will see, the DL is expected to be valid for distances in-between the Compton length of graviton $m^{-1}$, and a scale that is parametrically lower than the Vainshtein radius and can even reach the Schwarzschild radius. 

First, as should be clear from the discussion of section \ref{EXPNEW}, the DL should give a good description of the solution at least in the range $R_V \leq R \ll m^{-1}$. Another way to check this is to look more carefully at the expressions (\ref{SELFINTCAN}). Indeed, it is easy to check that the quadratic, cubic and quartic non derivative self interactions (\ref{SELFINTCAN}) (with appropriate values of $k_1, k_2, k_3, k_4$)  do not give significant corrections to the DL solutions for $R_V \leq R \ll m^{-1}$, while the cubic derivative $\tilde{\phi}$ self interaction is retained in the DL. Comparing the other interactions (\ref{SELFINTCAN}) to the kinetic terms in the regime where the linearized theory is expected to hold (i.e. for $ R \gg R_{V}$), one sees that those interactions have each  their own "Vainshtein" radii (corresponding to distances where those interactions generate ${\cal O}(1)$ correction to the linearized theory) which are strictly smaller than $R_V$, as first noted in Ref. \cite{Arkani-Hamed:2002sp}. Those radii are sent to zero in the DL, while $R_V$ is kept unchanged. 

Let us now discuss what is going on at distances smaller than $R_V$. Using again the expressions (\ref{SELFINTCAN}), it is easy to see that the quadratic, cubic and quartic non derivative self interactions stay negligeable at least up to $R_S$, while one can estimate when the other 
interactions become of the same order as the cubic interaction retained in the DL. Indeed, the Vainshtein scaling (\ref{nuSR}-\ref{muSR}) translates into the scaling $h \sim R_S/R$, $\partial \partial \phi \sim \mu \sim \sqrt{R_S/R}$ (via Eq. (\ref{PHIMU})), and $A \sim 0$. From which one easily obtains the scaling of the canonically normalized fields $\tilde{h}$, $\tilde{\phi}$, and $\tilde{A}$. Inserting this into 
Eq. (\ref{SELFINTCAN}) we see that a generic interaction (\ref{SELFINTCAN}) is much smaller than the cubic $\tilde{\phi}$
all the way down to the Schwarzschild radius $R_S$. Hence, as discussed in Ref. \cite{Creminelli:2005qk}, assuming the Vainshtein scaling, one expects the DL to be correctly describing the solution of the full nonlinear theory below $m^{-1}$ and down to $R_S$. The same question can be asked for the other scaling introduced above  in Eq. (\ref{muSRbis}). 
For this scaling, and below the Vainshtein radius, one still have $h \sim R_S/R$, while we have $\partial \partial \phi \sim \mu \sim m^2 R_V^4 R^{-2}$. To estimate the range of validity of the leading behaviour of the DL solution, one has to pay attention to the fact that the leading term in Eq. (\ref{muSRbis}) is a zero mode of the kinetic operator associated with the cubic term in the action of the DL. This kinetic operator, evaluated on this leading behaviour, is a sum of three terms which add to zero (see e.g. Eq. (\ref{eom mu})). Hence, one expects to see the solution leave the found DL behaviour whenever the left over  interactions (\ref{SELFINTCAN}) generates a term in the equation of motion larger than, or of the order of, any of those three terms. For a generic interaction  (\ref{SELFINTCAN}) this happens below the radius $R_{k_1,k_3,k_4}$ defined by
\ba
 R_{k_1,k_3,k_4} = R_V (R_S m)^{\frac{1}{5} + \frac{1}{5}\frac{3k_1+4 k_4}{k_1+ 2 k_3 -6}}.
\ea
Again we see that this scale is always parametrically smaller than the Vainshtein Radius, whenever $k_1+ 2 k_3 -6$ is stricly positive. When this later condition is not fulfilled, it is also easy to see that the neglected interactions are then always subdominant with respect to the cubic self interaction retained by the DL for $R \leq R_V$. Hence, for the scaling (\ref{muSRbis}), we expect as well the DL to give the leading behaviour of the solution of the full nonlinear system (if it exists) for distances below the Vainshtein radius and above the largest of the scales $ R_{k_1,k_3,k_4}$ which is simply 
$R_V \times a $. 

\subsection{Solving the decoupling limit equation of motion}
\label{results DL}

\subsubsection{Behaviour and boundary conditions at infinity} \label{421}
As we have just demonstrated, in the decoupling limit, one is left with the single nonlinear differential equation 
(\ref{Eqy1}) to be solved with appropriate boundary conditions. 
Note, that in the simple case of BD and AGS potentials (respectively potentials (\ref{S2}) and (\ref{S3})), equation (\ref{Eqy1}) (taking then the simpler form  (\ref{Eqysimp})) can be put in the well studied form (with obvious notation), encompassing in particular the six Painlev\'{e} transcendents 
\be
\ddot{w} = F(w,\xi) \dot{w}^2 + G(w,\xi) \dot{w} + H(w,\xi).\nonumber
\ee
However, one can easily check that the functions $F$, $G$, and $H$ above are such that the movable singularities of this equation are not only polar and even in this case solutions are not known analytically, hence one has to use a numerical integration in order to solve the equation of motion (\ref{Eqy1}).

This equation being of second order, we have to specify two initial conditions, $w(\xi_i)$ and $\dot{w}(\xi_i)$ at some point $\xi_i$ to start numerical integration from there. In our case, however, it is natural to start the integration from infinity, since 
we know the asymptotic behaviour of $w$ there.  Indeed, 
at infinity, i.e. large $\xi$ (or large $R$) regime, we are looking for a solution of equation (\ref{Eqy1}) which has the asymptotic behaviour given by dropping all the nonlinearities. 
Hence it should behave at large $\xi$ as 
\begin{equation} \label{largexi}
w(\xi) \sim w_\infty(\xi), 
\end{equation}
where $w_\infty(\xi)$ is defined by 
\begin{equation} \label{largexibis}
w_\infty(\xi) \equiv \frac{2 }{3 \xi^3}.
\end{equation}
This behaviour can then be used to integrate inward the equation of motion starting from the initial conditions  
\bea
\label{num cond}
w(\xi_i)&=&w_\infty(\xi_i), \\
\dot{w}(\xi_i)&=&\dot{w}_\infty(\xi_i).\nonumber
\eea
Taking a sufficiently large $\xi_i$, $\xi_i\gg 1$, we may hope that a 
solution we obtain by numerical integration will be close to the true solution (if it exists) of Eq.~(\ref{Eqy1})
with the asymptotic behaviour $w(\xi)\to w_\infty(\xi)$ as $\xi\to\infty$.

Things are slightly simpler in the case of equation (\ref{Eqysimp}) corresponding to the BD and AGS potentials (\ref{S2}) and (\ref{S3}). Indeed, in this case, the change of variable 
\ba
\Xii &=& \xi^{-3}, \\
\Wi &=&\left[w(\xi)\right]^{3/2},
\ea
puts the differential equation (\ref{Eqysimp}) in the form 
\be \label{wtnew}
s \Wi'' - \frac{\Wi^{1/3}}{4 \Xii^{8/3}}+ \frac{1}{6 \Xii^{5/3}\Wi^{1/3}}=0.
\ee
In this form, the first term in the left hand side above stands for all the term quadratic in $w$ in equation (\ref{Eqy1}) (i.e the terms in $Q$), 
the second term represent the term linear in $w$ and the last term  encodes the source. 
The boundary conditions (\ref{largexi}-\ref{largexibis}) translate into the Cauchy initial values at $0$
\ba 
\Wi(0)&=&0 \nonumber \\
\Wi'(0)&=& 0, \label{INITWT}
\ea 
while the asymptotic behaviour at $\Xii \rightarrow 0$  is obtained now by dropping the second derivative of $\Wi$ in the above equation (\ref{wtnew}) and reads 
\be
\Wi \sim \left(\frac{2}{3} \Xii \right)^{3/2}.\nonumber
\ee
In this simpler case, one sees clearly a crucial property of our problem, that holds also in the most general case (\ref{Eqy1}), 
namely the Cauchy problem (\ref{wtnew}-\ref{INITWT}) (or (\ref{Eqy1}) together with the asymptotic behaviour (\ref{largexi})) is a singular Cauchy problem, since the differential operator of equation (\ref{wtnew}) is obviously singular in $(\Wi,\Xii)=(0,0)$. 
As a result, one can not use standard theorems on Cauchy problem to conclude anything on the existence of a solution.  
Before turning to discuss solutions of our equation (\ref{Eqy1}), we first discuss in the following subsection a simple linear differential equation with a similar singular Cauchy problem to illustrate some crucial properties also found in our solving of equation (\ref{Eqy1}).

\subsubsection{A simple singular Cauchy problem} \label{422}
Consider the following  second-order differential equations:
\be
\label{example1}
y''(x)+y(x)=\frac{1}{x},
\ee
\be
\label{example2}
-y''(x)+y(x)=\frac{1}{x}.
\ee
These equations are singular at infinity similarly to Eq.~(\ref{Eqy1}). The term $y(x)$ and $1/x$ correspond respectively to the term linear in $w$ and the source term $1/\xi^3$, while the second derivative of $y$ can be related to the nonlinear differential operator $Q$. Aside from the fact the differential operator of equation (\ref{example1}) and (\ref{example2}) is linear, those equation have the same structure as equation (\ref{Eqy1}).
The solution of (\ref{example1}) is given by
\be
\label{exsolp}
y_1(x)=\bar{C}_1\cos(x)+\bar{C}_2\sin(x)+\text{Ci}(x)\sin(x)-\text{Si}(x)\cos(x),
\ee
and the solution of (\ref{example2}) is
\be
\label{exsolm}
y_2(x)=\bar{C}_1e^{x}+\bar{C}_2e^{-x}-\frac{1}{2}  \left( e^{ x }\text{Ei}(-x ) - e^{ -x }\text{Ei}(x)\right).
\ee
In Eq. (\ref{exsolp}) and (\ref{exsolm}), $\bar{C}_1$ and $\bar{C}_2$ are arbitrary constants, and 
$\text{Si}(x)$, $\text{Ci}(x)$ and $\text{Ei}(x)$ are sine integral, cosine integral and exponential integral
functions correspondingly, given by
\be
\text{Si}(x) = \int_0^x\frac{\sin t}{t}dt, \quad 
\text{Ci}(x) = -\int_x^\infty\frac{\cos t}{t}dt, \quad
\text{Ei}(x) = -\int_{-x}^\infty\frac{\exp(- t)}{t}dt.\nonumber
\ee
Now let us find solution(s) of (\ref{example1}) and (\ref{example2}) such that,
\be
\label{excond}
y_{1,2}(x)\to \frac{1}{x}, \; \text{when}\; x\to\infty.
\ee
This correspond, {\it mutatis mutandis}, to the asymptotic at infinity, Eq. (\ref{largexibis}).
There is a unique solution of (\ref{example1}) with the asymptotic (\ref{excond}),
\be
\label{ex1sol}
y_1(x)=\frac{\pi}{2}\cos(x)+\text{Ci}(x)\sin(x)-\text{Si}(x)\cos(x),
\ee
while for the equation (\ref{example2}) the asymptotic behaviour (\ref{excond}) does not
fix uniquely the solution,
\be
\label{ex2sol}
y_2(x)=\bar{C}_2e^{-x}-\frac{1}{2}  \left( e^{ x }\text{Ei}(-x ) - e^{ -x }\text{Ei}(x)\right).
\ee
I.e. one has the freedom to choose at will the integration constant $\bar{C}_2$.
Solution (\ref{ex1sol}) has the following asymptotic expansion
\be
\label{ex1e}
y_1(x)=\frac{1}{x}-\frac{2}{x^3}+\frac{24}{x^5}-\frac{720}{x^7}+O\left(\frac{1}{x^9}\right),\nonumber
\ee
while looking for a power serie expansion of solution (\ref{ex2sol}), we find 
\bea \nonumber
y_2(x) &=& \frac{1}{x}+\frac{2}{x^3}+\frac{24}{x^5}+\frac{720}{x^7}+O\left(\frac{1}{x^9}\right).
\eea
Those series expansions are in fact divergent, and hence only give at best asymptotic expansions of the solution, however we notice that the last one misses the existence of the homogeneous mode $\bar{C}_2e^{-x}$.
One can nonetheless easily see that the solutions (\ref{ex1sol}) and (\ref{ex2sol}) 
are in fact finite and they give the correct asymptotic (\ref{excond}) at the infinity.

The toy examples (\ref{example1}) and (\ref{example2}) show that for  singular equations, the asymptotic behaviour 
at infinity can be enough to fix uniquely the solution (and hence the singular Cauchy problem is well-posed), this is the case for example of Eq. (\ref{example1}); while in other cases, such as that of Eq. (\ref{example2}), there exist a family of solutions with the same asymptotic behaviour, which can be missed by looking for the solution by a power serie expansion. The properties summarized in this last paragraph are exactly recovered for our equation (\ref{Eqy1}), as we now see, beginning by discussing the simplest case of BD and AGS potentials.  

\subsubsection{Solutions for the BD and AGS potentials outside the source} \label{BDAGSout}
\begin{figure}[h!]
\begin{center}
\includegraphics{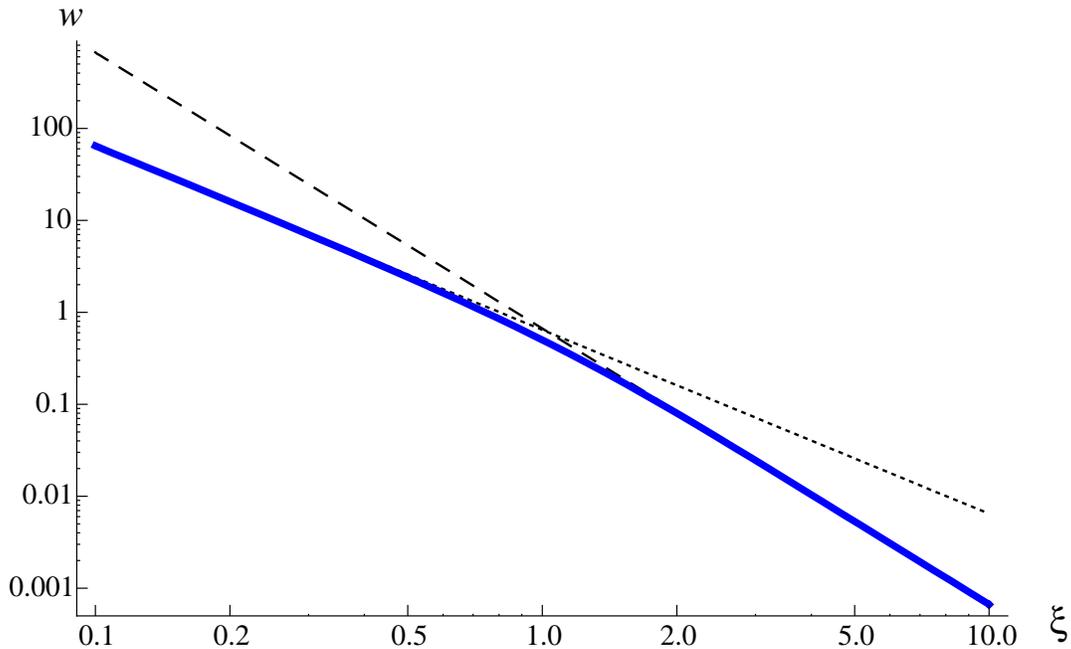}
\caption{Solution for $w$ in the case of the \BD\ potential. The numerical 
solution  is shown by solid thick (blue) line. 
For distances much larger than the Vainshtein radius, $\xi\gg 1$,  the solution is well approximated by 
the asymptotic (\ref{largexibis}), shown by dashed thin (black) line.  
Close to the source, $\xi\ll 1$, the numerical solution approaches the $Q$-scaling asymptotic, Eq.~(\ref{QSCALING}),
shown by dotted thin (black) line.}
\label{plotmuBD}
\end{center}
\end{figure}

\begin{figure}[h!]
\begin{center}
\includegraphics{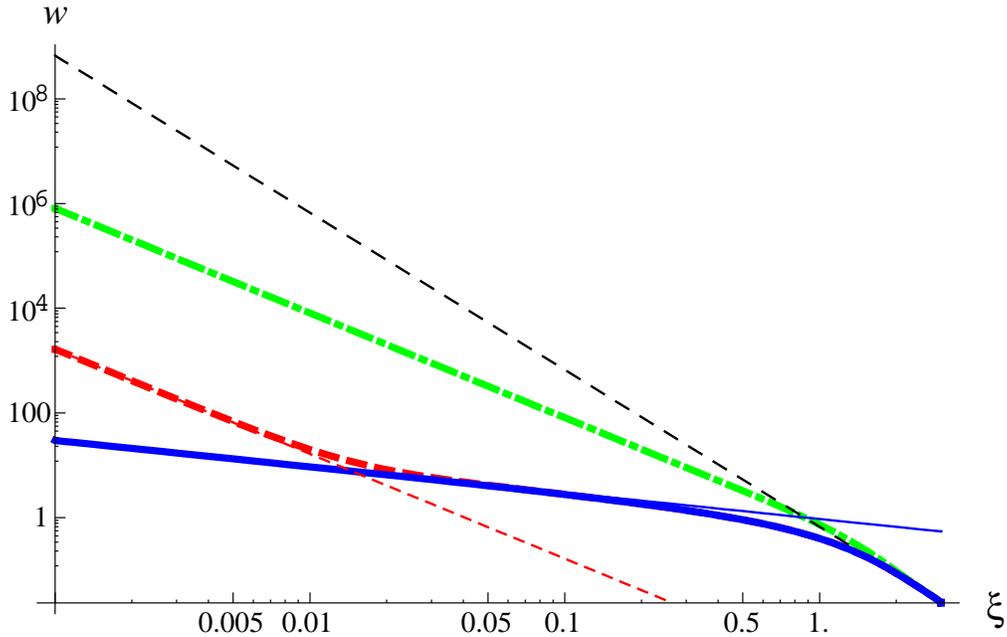}
\caption{Plot of the numerical solutions for $w$ in the case of the \AGS\ potential, starting from $\xi_{i}=2.5$. 
For $\xi\gg 1$ all the solutions approach the asymptotic (\ref{largexibis}), shown by dashed thin (black) line. 
At small distances, $\xi\ll 1$ the solutions pick up different asymptotic regimes.
The solid thick (blue) line corresponds to the Vainshtein scaling $w\sim 1/\sqrt{\xi}$.
The dash-dotted thick (green) line corresponds to a solution with the $Q$-scaling (\ref{QSCALING}).
The dashed thick (red) line corresponds to a solution which first follows 
the Vainshtein scaling and then finally picks up the $Q$-scaling (\ref{QSCALING}).
}
\label{AGSplot}
\end{center}
\end{figure}

In the case of the BD and AGS potentials (of Eq.~(\ref{S2}) and (\ref{S3}) respectively), the nonlinear differential equation to solve takes the simple form (\ref{Eqysimp}). 
In those cases, it is easy to find a power serie expansion of the solution around $\xi=+\infty$ in the form 
\be\label{ASYMEXP}
w(\xi)=\sum_{n=0}^{\infty}\frac{w_{n}}{\xi^{3+5n}}\;.
\ee
The first non trivial six $w_{n}$ coefficients, are given by 
\be \label{ASYMEXPCOEF}
w(\xi)=\frac{2}{3\xi^{3}}+ s \frac{4}{3\xi^{8}}+\frac{1024}{27\xi^{13}}+ s \frac{712960}{243\;\xi^{18}}+\frac{104910848}{243 \;\xi^{23}} + s \frac{225030664192}{2187 \;\xi^{28}}+...
\ee
Note that the first two terms of this series match those given in equation (\ref{muLR}), and in fact this expansion is just the continuation of the one given in this equation. One can check numerically that this power series is in fact divergent. However we found it useful for the purpose of numerical integration to use this expansion as an asymptotic expansion of the solutions. Indeed it provides a better approximation of the solution than keeping only the leading order (\ref{num cond}), provided that we truncate the expansion at the appropriate order \footnote{I.e., as usual with asymptotic expansions, for any given $\xi_i$ from which we want to integrate inwards, 
there is an order $n$ that minimizes the $n$-th term ${w_{n}}/{\xi^{3+5n}}$ of the series (\ref{ASYMEXP}), and we found it helpful to use the formal series truncated at this order $n^{(\xi)}$, i.e. to replace the initial conditions (\ref{num cond}) by
\ba
w(\xi_i)&=&\sum_{k=0}^{n^{(\xi_{i})}}\frac{w_{k}}{\xi^{3+5k}}\;,\nonumber  \\
\dot{w}(\xi_i)&=&\frac{d}{d \xi}\left(\sum_{k=0}^{n^{(\xi_{i})}}\frac{w_{k}}{\xi^{3+5k}}\right)\label{num cond expansion}.
\ea}.
In both cases of BD and AGS potentials, we were able to integrate numerically inwards equation (\ref{Eqysimp}) 
 and obtain non singular solutions all the way to small radii. Our results are shown in figures \ref{plotmuBD} and \ref{AGSplot}. 
 
These numerical solutions exhibit different behaviours at small $\xi$ which can be understood as we now explain. First, if, following Vainshtein, we assume that we can drop  the term linear in $w$ in equation 
(\ref{Eqysimp}), the leading behaviour should be given by a solution of the equation 
\be \label{QW0}
 2 Q(w)=\frac{1}{\xi^{3}}, \nonumber
\ee 
namely the same equation as equation (\ref{EQQSIMP}). 
An assumed power law behaviour leads then to the {\it Vainshtein} scaling at small $\xi$, as (see Eq. (\ref{muSR}))
\be \label{VAINSCA}
w \sim w_{V} \propto \frac{1}{\sqrt{\xi}}.
\ee  
Whether such a scaling leads to a real or an imaginary solution depends on the potential.
For the \AGS\ potential it leads to a real leading behaviour $w \sim w_{V}=\sqrt{\frac{8}{9\; \xi}}$. 
An expansion of the solution around this Vainshtein scaling can then be obtained in the following form 
\be\label{Vainshtein generalized expansion}
\begin{aligned}
w(\xi)=\sqrt{\frac{8}{9\; \xi}}+B_{0} \; \xi^{-\frac{5}{4}+\frac{3 \sqrt{5}}{4}}-\frac{3 \left(-5+\sqrt{5}\right) }{8 \sqrt{2} \left(-4+\sqrt{5}\right)}B_{0}^2 \;\xi^{-2+\frac{3 \sqrt{5}}{2}}+\frac{6 \xi^2}{31}+...
\end{aligned}
\ee
This expansion can be seen to match correctly the (only - see thereafter) solution of the AGS case that has the small $\xi$ Vainshtein asymptotics, provided the constant $B_0$ is fixed to some specific value. It can be understood as follows. The leading term is of course the Vainshtein scaling. The next term $\propto  \xi^{\frac{1}{4}\left(-5+3\sqrt{5}\right)}$ is a zero mode of the linearized $Q$ around the leading behaviour. The rest of the series is then obtained, as usual, order by order in $\xi$.

\begin{figure}[h!]
\begin{center}
\includegraphics{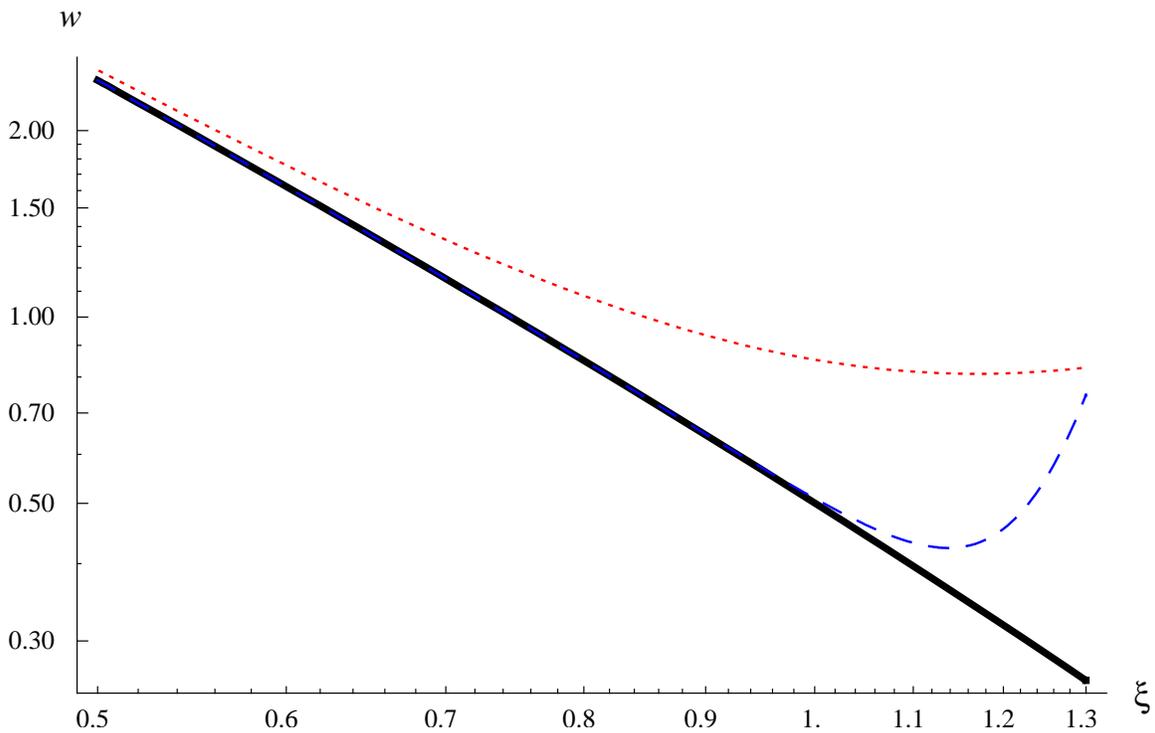}
\caption{Plot of the numerical solution (solid black curve), and the series expansion (\ref{w-DL}) given up to order $\xi$ (dotted red curve) and up to order $\xi^{13}$ (dashed blue curve). The expansion up to $\xi^{13}$ approximates the numerical solution with a precision better that $99\%$ in the range $0\leqslant \xi \leqslant 0.95$.}
\label{plotDLwSmallXi}
\end{center}
\end{figure}

For the BD potential, however, the Vainshtein scaling at small $\xi$ leads  to an imaginary solution, which is not acceptable. In the BD case, rather than the Vainshtein scaling, we found that there exists a real non singular numerical solution all the way to small $\xi$ 
which interpolates between the large distance behaviour $w\sim 1/\xi^{3}$ for $\xi\gg1$ and the small distance behaviour 
$w\sim 1/\xi^{2}$ for $\xi\ll1$ (see Figure \ref{plotmuBD}). Notice that this solution is perfectly regular around the Vainshtein radius $\xi_V=1$. 
The found new scaling at small $\xi$, namely 
\be \label{QSCALING}
w\sim 1/\xi^{2},
\ee
and named in the following {\it $Q$-scaling}, is in fact obtained from a zero mode of the operator $Q$. Indeed one can check that the equation $Q(w)=0$ has the exact solution\footnote{This solution can be easily found  using the new function and variables 
\bea
\xi &\equiv& \Xio^{1/3},\nonumber\\
w(\xi) &\equiv& \left(\frac{\Wo(\Xio)}{\Xio}\right)^{2/3}\;,\nonumber
\eea
which translates the differential equation (\ref{Eqysimp}) into 
the following nonlinear equation on $\Wo(\Xio)$
\be
\label{yeq}
s\frac{d^{2}\Wo(t)}{d\Xio^{2}}-\frac{\Wo^{1/3}}{4\Xio^{2/3}}+\frac{1}{6\Xio \Wo^{1/3}}=0, \nonumber
\ee
where $Q$ is transformed into the first term on the left hand side.}
\ba\label{Q scaling}
w_{Q}(\xi)&=&\left(\frac{K_{0}}{\xi^{3}}+K_{1}\right)^{\frac{2}{3}} \label{wQ2bis}\nonumber\\
&=&\frac{A_{0}}{\xi^{2}}+B_{0}\; \xi+{\cal O}\left(\xi^4\right)\label{wQ2},\nonumber
\ea
where $K_{0}$ and $K_{1}$ are two arbitrary constants and $A_{0}=K_{0}^{2/3}, B_{0}=(2K_{1})/(3K_{0}^{1/3})$ are the first two coefficients of the expansion of the zero mode around $\xi=0$. 
A solution of equation (\ref{Eqysimp}) can then be found in the form of the double expansion
\be\label{SMALLXIEXP}
w(\xi)=\frac{A_{0}}{\xi^{2}}+\sum_{n=1}^{\infty}\sum_{k=0}^{n}w_{n,k}\; \xi^{n}(\ln \xi)^{k}
\ee 
where the coefficients $w_{n,k}$ are functions of $A_{0},B_{0}$. For the \BD\ and \AGS\ potentials, the first coefficients read:
\be
\begin{aligned}
w(\xi)=&\frac{A_0}{\xi^2}+\frac{3A_{0}B_0 -s \;\ln
   \xi}{3 A_0}\; \xi + s \; \frac{3}{8}\;  \xi^2\\
  & +\frac{1 + s \;6 A_0
   B_0-54{A_0}^{2} B_0^{2}-\left(2 -s \;36
   {A_0} B_0\right) \ln \xi-6 \ln ^2\xi}{216
   A_0^3}\;\xi^4+O\left(\xi^5\right)\; . \label{w-DL}
\end{aligned}
\ee
The values of the parameters $A_{0},B_{0}$ can be fixed matching numerically the expansion at small $\xi$ with the large $\xi$ asymptotic behaviour. In the case of the BD potential, we find $A_{0} \sim  0.645$ and $B_{0} \sim 0.208$
and we find that the expansion fits very well the numerical solution up to  $\x\sim 1$ (the Vainshtein radius), as can be seen in Fig.~\ref{plotDLwSmallXi}.
Expansions for $u(\xi)$ and $v(\xi)$ can be found in the same way. They read
\ba\label{u DL simplest case}
u(\xi)&=&\frac{1}{\xi}-\frac{A_{0}}{2}- \frac{\xi^3}{6A_{0}} \left(3 A_{0} B_{0} -s \; \ln\xi\right) - \frac{3 s\; \xi^4}{16}+O\left(\x^5\right)\label{u-DL} \\
v(\xi)&=&-\frac{1}{\xi}+(D_0+\frac{1}{2} A_{0} \ln\xi)\nonumber\\
&&\hspace{1.5cm}+\frac{\xi^{3}}{54 A_{0}}\left(s  + 9{A_{0}}B_{0} -s \; 3\ln\xi \right) +s \; \frac{3\xi^{4}}{64}+O\left(\x^5\right)\label{v-DL}\label{v DL simplest case}.
\ea
In dimensionful units, and keeping only the dominant terms, the above equations (\ref{u DL simplest case}) and (\ref{v DL simplest case}) read (the corresponding expression for $\mu$ has been given in equation (\ref{muSRbis}))
\ba
\lambda&=&\frac{R_{S}}{R}+\mathcal{O}(1)\\
\nu&=&-\frac{R_{S}}{R}+\frac{1}{2} A_{0} (R_{S} M)^{4/5}\ln R+\mathcal{O}(1) \label{nuQsca}
\ea
Note that the leading terms at small $R$, $\lambda=R_{S}/R$ and $\nu=-R_{S}/R$, correspond to the linearization of the Schwarzschild solution in General Relativity, and hence, even with our new  scaling, one recovers (in the decoupling limit) 
the Schwarzschild solution at distances smaller than the Vainshtein radius, in agreement with the Vainshtein original idea that nonlinearities can cure the vDVZ discontinuity. Notice however that this happens here for a potential, the BD potential, for which the Vainshtein mechanism was believed not to work because of the imaginary nature of the Vainshtein scaling at small $\xi$ \cite{Damour:2002gp}. It is also interesting to notice that the first correction to the Newtonian potential arising from equation (\ref{nuQsca}) has a form similar to the one required by MOND \cite{Milgrom:1983ca}. 

Our numerical investigations of the solution in the BD case lead to the conclusion that there is in this case a unique solution 
(shown in figure \ref{plotmuBD}) with the right asymptotics (\ref{largexibis}). This was confirmed by an analytic proof of this uniqueness (around $\xi = +\infty$) provided by J.~Ecalle \cite{Ecalle}. To conclude on the BD potential, let us further mention that in this case, the numerical integration appears to be stable against small perturbations. Namely, a slight change in the initial conditions at large $\xi_i$ does not result in growing divergences but only in small oscillations around the found unique solution. This can be understood analytically by perturbing around the asymptotic solution (see appendix \ref{LINEARPERT}). 

Let us now turn to discuss the AGS case (potential (\ref{S3}), or $s=+1$). In this case, the expansion 
(\ref{ASYMEXPCOEF}) still holds, but in analogy with the example (\ref{example2}) we found in this case infinitely many solutions with the same power serie expansion (\ref{ASYMEXPCOEF}) at infinity. In this case, indeed, we found that with the same asymptotics at infinity, different behaviours, including the original Vainshtein scaling, were possible at small $\xi$.
In fact, the most general behaviour at small $\xi$, is still of the form of the $Q$-scaling (\ref{QSCALING}) with the asymptotic expansion (\ref{SMALLXIEXP}-\ref{w-DL}), but depending on the chosen solution at infinity, this behaviour can be picked up at arbitrarily small $\xi$ while there is an intermediate region at $\xi$ smaller than one where the solution follows the Vainshtein scaling (\ref{VAINSCA}) (see Fig. \ref{AGSplot}). The Vainshtein scaling then appears as the limiting case of the family of solution having the right asymptotics as infinity (\ref{largexibis}), and so can be obtained in the AGS case.
From a practical point of view, numerical integration in the \AGS\ case is quite difficult, due to numerical instabilities which appear for distances larger that the Vainshtein radius $\xi=1$. Indeed, any small departure from the solution sources exponentially growing modes and Runge-Kutta type of integrations reach singularities quickly. 
In this context, we found very useful, while integrating from large $\xi_{i}$ towards $\xi=0$, to use the asymptotic expansion (\ref{ASYMEXP}), truncated at the order appropriate to $\xi_{i}$. We were also able to confirm our numerical results by solving  equation (\ref{Eqy1}) using a relaxation method, which is more robust under numerical instabilities.
 Again, our numerical investigations were confirmed by mathematical proofs of the existence (around $\xi = + \infty$) of infinitely many solutions with asymptotics  
(\ref{largexibis}) \cite{Ecalle,SPAIN}. In fact, one can also show that any two solutions of (\ref{Eqy1}) having the same asymptotics (\ref{largexibis}) differ (at dominant order) at large $\xi$ by a quantity given by 
a constant times $\xi^{3/2} \exp\left(- k \; 3/5 \; \xi^{5/2}\right)$ where $k$ is an integer \cite{Ecalle}.  Finally, one can also analyse the problem by perturbation theory around the 
large $\xi$ behaviour (\ref{largexibis}). This is done in appendix \ref{LINEARPERT} and we find that there are both a growing and a decaying mode. The growing mode is responsible for the numerical instability pointed out above and should be discarded in the true solution; the subdominant decaying mode, similar to the decreasing exponential found for equation (\ref{example2}), can be freely specified while keeping the asymptotic behaviour (\ref{largexibis}).

\subsubsection{Solution for the BD and AGS potentials inside the source}\label{BDAGSin}

\begin{figure}[h!]
\begin{center}
\includegraphics{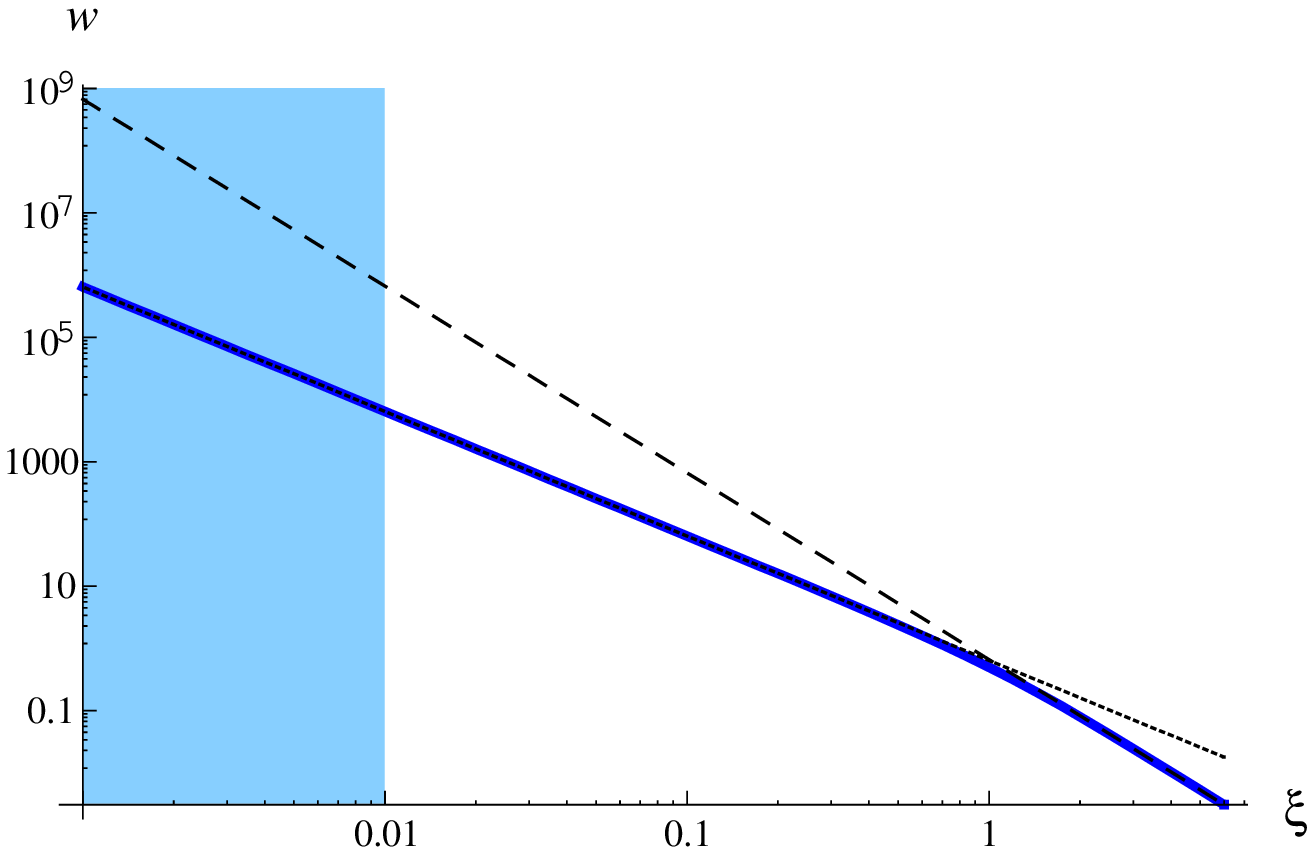}
\caption{Plot of the numerical solution for $w$ in the case of the \BD\ potential in the presence of a source of radius $\xi_{\odot}=0.01$ (depicted above by a light blue area), starting from $\xi_{i}=6$. The blue thick line corresponds to the  $Q$-scaling of Eq. (\ref{diverging solution at small distance}), which is picked up by the asymptotic behaviour at infinity. Note that the solution is almost not affected by the presence of the source.}
\label{BDplotSource}
\end{center}
\end{figure}

\begin{figure}[h!]
\begin{center}
\includegraphics{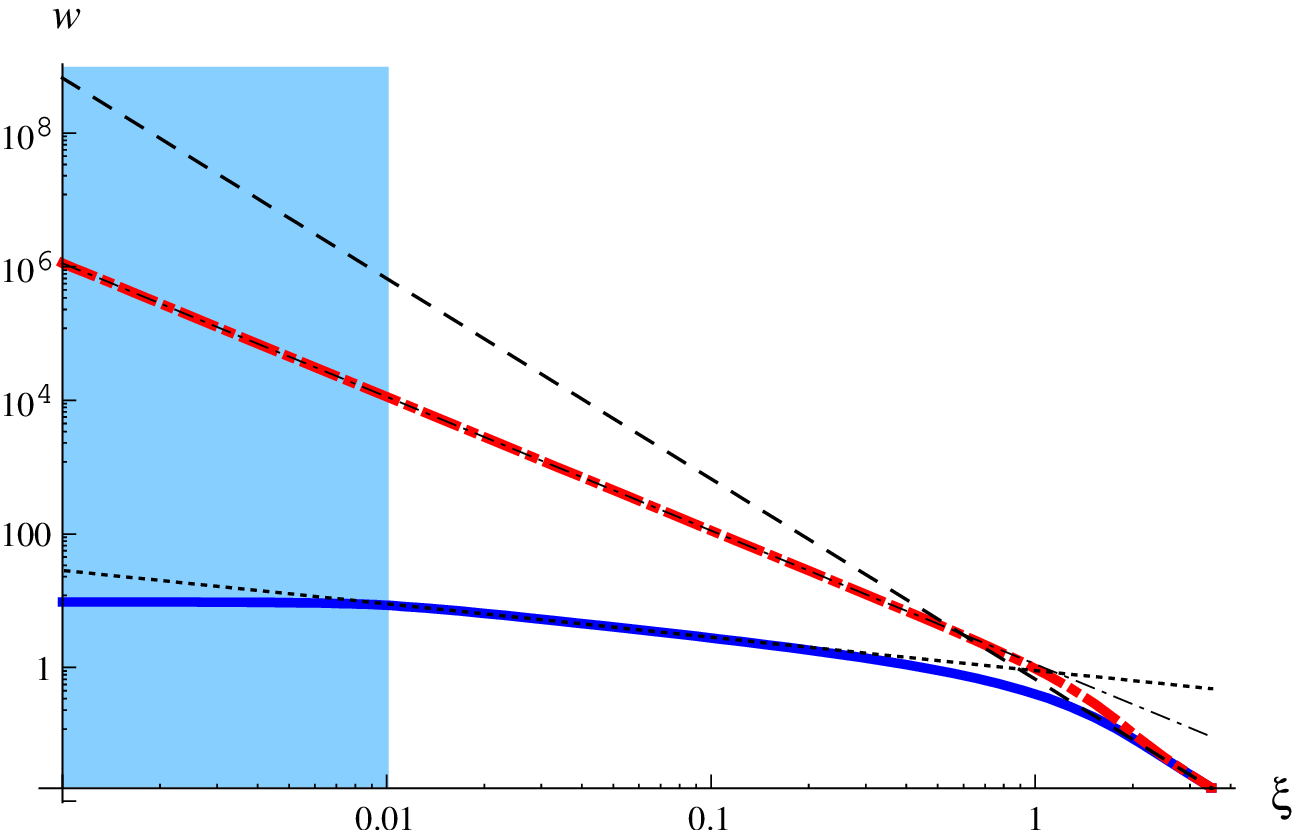}
\caption{Plot of the numerical solutions for $w$ in the case of the \AGS\ potential in the presence of a source of radius 
$\xi_{\odot}=0.01$ (depicted above by a light blue area), starting from $\xi_{i}=3.5$. 
The solid thick (blue) line corresponds to the solution having the 
behaviour Eq. (\ref{constant solution at small distance}) inside the source.
The thick dash-dotted (red) line illustrates the $Q$-scaling, Eq. (\ref{diverging solution at small distance}), inside the source.}
\label{AGSplotSource}
\end{center}
\end{figure}

Let's now include a source. For simplicity, we consider a star of constant density, $\rho_a$, and radius $\xi_{\odot}$ (i.e. of radius $R_{\odot}$ in physical units). We have 
\be
\rho_{a}=\frac{3}{\xi_{\odot}^{3}}, \nonumber
\ee
leading to the equation for $w$ inside the source
\be
2Q(w)+\frac{3}{2}w=\frac{1}{\xi_{\odot}^{3}}\; .\nonumber
\ee
Here too, two different types of solutions can be guessed. The first kind are solutions of the form
\be\label{diverging solution at small distance}
w(\xi)=\frac{A_{0}}{\xi^{2}}+\sum_{n=1}^{\infty}w_{n}\xi^{n}
\ee
corresponding again to a $1/\xi^{2}$ leading behaviour for small $\xi$. The second kind are solutions of the form 
\be\label{constant solution at small distance}
w(\xi)=\sum_{n=0}^{\infty}w_{2n}\xi^{2n}=A_{0}+s\frac{3A_{0}\xi_{\odot}^{3}-2}{20A_{0}\xi_{\odot}^{3}}\;\xi^{2}+...
\ee
where the leading term is now $w\sim A_{0}$  (and $s=-1$ in the \BD\ case, and  $s=+1$ in the \AGS\ one).

Our numerical integration indicates that, the \BD\ solution, which is fixed by the asymptotic condition at infinity (\ref{num cond}), follows  the $w\sim 1/\xi^{2}$ behaviour inside the star, as shown in Fig. \ref{BDplotSource}. In the \AGS\ case, the situation is more subtle: generically (i.e. without fine tuning), the non singular solutions also adopt such a $1/\xi^{2}$ leading behaviour; an example of such a solution is presented in Fig. \ref{AGSplotSource} (thick dash-dotted red curve). However, there is a case for which the other scaling $w\sim A_{0}$ is possible. This case precisely corresponds to the Vainshtein solution outside of the source, as shown in Fig. \ref{AGSplotSource} (thick blue curve). 

Note that for the Vainshtein and Q-scaling at small $R$, and in the absence of a source, the Jacobian (\ref{JACOB}) is going asymptotically to zero. This is not true for the scaling (\ref{constant solution at small distance}). Moreover, only this scaling leads to a finite Ricci scalar of the physical metric $g_{\mu\nu}$ at $R=0$. Hence, this might provide a way to select a physical solution in the full nonlinear theory. Discussion of these interesting issues is however left for a future work \cite{future}, since it involves keeping track of nonlinear terms in the equations beyond those appearing in the DL.  
\label{424}

\subsubsection{Solution in the most general case}\label{425}
In the most general case of equation (\ref{Eqy1}) with arbitrary $\alpha$ and $\beta$, a solution 
with a Vainshtein scaling at small $\xi$ can easily be found. It is obtained by a power law ansatz and dropping the term linear in $w$ in Eq.(\ref{Eqy1}) and reads 
\be \label{VASCA}
w_V(\xi)=\left[\frac{16}{3(25\alpha+13\beta)}\right]^{1/2}\xi^{-1/2},\quad {\rm as} \quad \xi\to 0.\nonumber
\ee
This solution is real iff 
\be
25\alpha+13\beta>0.\nonumber
\ee
In analogy with the discussion of the previous subsection, one can look for another scaling where the dominant term (at small $\xi$) is given by a zero mode of the operator $Q$ appearing in equation (\ref{Eqy1}).
Zero modes in the form of power law,
\be 
w_Q=A \xi^p, \label{zeromode}
\ee
can be found, where $p$ is given by 
\bea
\label{p12}
p_{\pm}&=&\frac{-3\alpha-2\beta \pm \sqrt{-\beta^2-2\alpha\beta}}{\alpha+\beta}, 
\quad {\rm for}\quad \alpha+\beta\neq 0,
\eea
where the plus sign in the right hand side corresponds to the power $p_+$, and the minus sign to the power $p_-$. Note that 
for $\alpha+\beta=0$, e.g. the BD and AGS potentials studied in subsection \ref{BDAGSout} and \ref{BDAGSin}, the $Q$-scaling 
$p=-2$ can 
be found from (\ref{p12}) taking the limit $\alpha\to -\beta$. One can check also that, in order to describe the dominant behaviour at small $\xi$, any particular root, $p_+$ or $p_-$, should satisfy an additional criterion,
namely,
\be
\label{p12crit}
p<-\frac12,\nonumber
\ee
otherwise with this scaling the nonlinear terms in Eq.~(\ref{Eqy1}) are not dominant.

\begin{figure}[t]
\begin{center}
\includegraphics[width=0.8\textwidth]{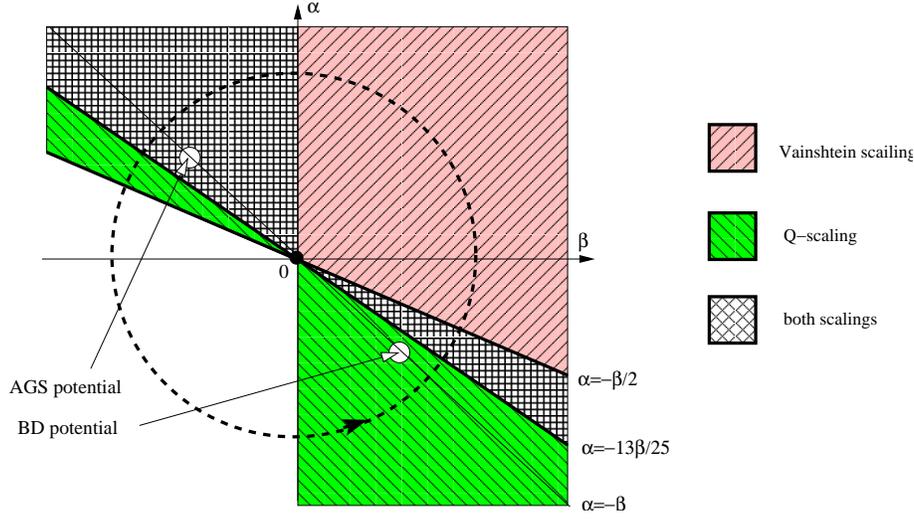}
\end{center}
\caption{\label{figsc} The diagram of possible scalings for solutions of Eq.~(\ref{Eqy1}) at $\xi\to 0$ 
in the plane of parameters $\alpha$ and $\beta$. The whole parameter  space (excluding the origin) can be reduced to the circle of unit radius, with the help of an appropriate rescaling (see the text).}
\end{figure}

\begin{figure}[t]
\psfrag{inf1}[l]{\tiny{$-\infty$}}
\psfrag{inf2}[l]{\tiny{$+\infty$}}
\psfrag{inf3}[l]{\tiny{$-\infty$}}
\begin{center}
\includegraphics[width=\textwidth]{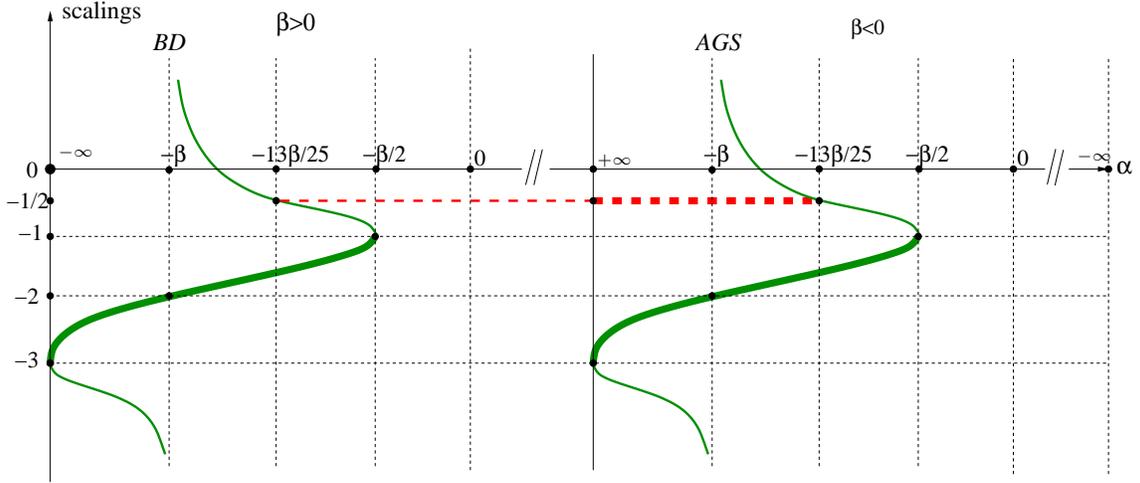}
\end{center}
\caption{\label{figab} The values of the possible power law scaling exponents of the solutions of Eq.~(\ref{Eqy}) at small $\xi$
as a function of parameters $\alpha$ and $\beta$. The circle in the Fig.~\ref{figsc} is mapped to the horizontal 
segment. The Vainshtein scaling is shown by dashed horizontal (red) line, 
while the $Q$-scalings are shown by solid (green) lines.
The scalings which are found to smoothly continue to  the asymptotic (\ref{largexibis})
at large $\xi$ are shown by thick lines. The scalings at $\xi\to 0$ which do not give the required 
asymptotic at large $\xi$ are shown by thin lines.}
\end{figure}

The various possibilities for the different power law scaling as $\xi\to 0$ are summarized in  Fig.~\ref{figsc} in the $(\alpha,\beta)$ plane. Roughly speaking Eq.~(\ref{Eqy1}) possesses 
four different regimes at $\xi\to 0$ with:
\begin{itemize}
\item only Vainshtein scaling; 
\item only $Q$-scaling;
\item both Vainshtein and $Q$-scalings;
\item no power law scaling.
\end{itemize}
There are three boundary lines, where interesting 
transitions from one type to another can happen:
\begin{itemize}
\item $\beta=0$; 
\item $\alpha=-13\beta/25$;
\item $\alpha=-\beta/2$.
\end{itemize}
Note, that with the help 
of the rescaling $\xi\to \bar{C} \xi$ and  $w\to \bar{C}^{-3} w$, with $\bar{C}= \left(\alpha^2+\beta^2\right)^{1/10}$,  
Eq.~(\ref{Eqy1})  can be brought to the same  
form where the coefficients $\alpha$ and $\beta$ now verify $\alpha^2+\beta^2=1$, and hence one can restrict the 
study of the different cases in the $(\alpha,\beta)$ plane of Fig. \ref{figsc} to those where $\alpha$ and $\beta$ lie on the circle of unit radius. This circle is mapped to the horizontal segment in Fig.~\ref{figab} which shows together with the possible power scaling (the values of the Vainshtein and $Q$-scaling power scalings are plotted on the vertical axis), results of the numerical integrations of equation (\ref{Eqy1}) indicating which of the scalings are in fact realized in the solutions having the right large $\xi$ asymptotics $w_{\infty}$ given by (\ref{largexibis}).
Depending on the sign of $\beta$, this asymptotics does or does not fix a unique behaviour at small $\xi$. Indeed, 
for $\beta>0$ small variations of initial conditions at large $\xi$ does not affect the solution at small $\xi$,
similar to the case we have already studied, the BD potential. This signals about the uniqueness of the solution 
for  Eq.~(\ref{Eqy1}) when the asymptotic behaviour is fixed by (\ref{largexibis}).
On the contrary, for $\beta<0$, the asymptotic behaviour (\ref{largexibis}) does not fix uniquely the solution,
and we find infinitely many different solutions at small $\xi$  with the same 
behaviour at large $\xi$. 

Starting from the point $(\alpha,\beta)=(-1,0)$, and moving counter-counter-clock-wise along the
unit circle of figure \ref{figsc}, i.e. from the left and along the horizontal segment of figure \ref{figab}, let us now describe with more details the successive cases of interest together with the results of our numerical integrations.

\begin{itemize}
\item {$\beta=0,\, \alpha<0$}. 

This point is somewhat special: 
no Vainshtein scaling is possible and  the two $Q$-scalings are equal and coincide 
with the asymptotic solution at infinity,
\be
w_Q(\xi)\propto \xi^{-3}, \quad \xi\to 0.\nonumber
\ee
In fact, in this case the asymptotic form Eq.~(\ref{largexibis})
satisfies Eq.~(\ref{Eqy1}) exactly for any $\xi>0$,
\be
w(\xi)=\frac{2}{3\xi^3}.\nonumber
\ee
This can be easily understood recalling that whenever $\beta$ vanishes, the covariant form of $Q^{(\alpha,\beta)}$ can be read off in equation (\ref{tildephi}) to contain only $\Box$ acting on $\tilde{\phi}$, and hence, in vacuum, a solution of $\Box \tilde{\phi} =0$ is a solution everywhere\footnote{We leave for a future publication the discussion of the inclusion of a source.}.

\item {$\beta>0, \,-\infty<\alpha\leq -\beta/2$.} 

The solution at $\xi\to 0$, with the correct large distance asymptotics (\ref{largexibis}), is always given by the $Q$-scaling with $p_-$, although 
in some parts of this region of parameters we could expect the second $Q$-scaling and/or Vainshtein 
scaling. This region contains the BD potential, which we have already studied in detail. 

\item {$\beta>0, \,-\beta/2<\alpha< +\infty$.} 

The solution is always singular in this region, the integration from large $\xi$ always breaks down at 
some finite distance $\xi>0$. One could have expected the existence of  a regular solution 
with the Vainshtein asymptotic scaling but it appears that this scaling does not give the required asymptotic behaviour at infinity.

\item {$\beta=0,\, \alpha>0$}. 

This case is similar to the case, $\beta=0,\, \alpha<0$. An exact solution for $\xi>0$ is given by an 
analytic expression,
\be
w(\xi)=\frac{2}{3\xi^3}.\nonumber
\ee

\item {$\beta<0, \,-13\beta/25<\alpha< +\infty$.} 

For $\beta<0$ the asymptotic condition at infinity (\ref{largexibis}) does not fix uniquely the solution, as
in the case of AGS potential, which is included in this region of parameters. Thus tuning the initial conditions
at large $\xi$ we are able to obtain at small $\xi$ either the $Q$-scaling with $p_+$\footnote{Note that this is 
a whole family of solutions, since  by tuning the initial conditions at large $\xi$ we can get different factor $A$ in 
(\ref{zeromode}) at small $\xi$.} 
or the Vainshtein scaling or a singular solution.
It is remarkable that the other $Q$-scaling cannot be obtained by the tuning of the initial conditions
 even though the scaling with $p_-$ would be a dominant scaling for $-\beta<\alpha< +\infty$.
Here again it is found, as in the AGS case, that the Vainshtein scaling appears as a limiting regime of a family of solutions which all have the $Q$-scaling as their innermost scaling, but can behave at some intermediate range of $\xi$ following the Vainshtein scaling (see figure \ref{AGSplot}).

\item {$\beta<0, \,   -\beta/2 \leq \alpha\leq -13\beta/25$.} 

The only possible scaling here, as confirmed by numerical integration\footnote{the border case with $p_+=p_-$ is included.}, is a family of the $Q$-scaling with $p_+$ and different $A$ in (\ref{zeromode}).  
Of course, changes in the initial conditions also lead to singular solutions as well.

\item {$\beta<0, \,   -\infty <\alpha< -\beta/2$.} 

There is no scaling possible for small $\xi$, and indeed, the numerical integration gives only 
singular solutions in this region. 

\end{itemize}
When it is possible to find a scaling for small $\xi$ (either of Vainshtein or of $Q$ type), a full solution can be obtained as a power law expansion in $\xi$, similar to the ones obtained in the specific cases of  \AGS\  and \BD\ potentials (Eq.(\ref{Vainshtein generalized expansion}) and (\ref{w-DL})). Note that for both Vainshtein and $Q$-scalings, it is always possible to add to the leading behaviour a zero mode of the linearized $Q$ around the leading term, introducing a free constant (denoted $B_{0}$ in Eq. (\ref{Vainshtein generalized expansion}) and (\ref{w-DL})) which should be fixed by matching the small $\xi$ expansion with the asymptotic behaviour  (\ref{largexibis}).

To conclude, 
we have found that for the range of parameters 
$\beta<0,\, \alpha<-\beta/2$ and $\beta>0,\, \alpha>-\beta/2$,  regular solution(s) with the right large distance asymptotics (\ref{largexibis}) exist. For positive $\beta$, and fixed $\alpha$ only one such solution exists and it has the $Q$-scaling at small $\xi$. For negative $\beta$, and fixed $\alpha$, a whole family of such solutions exist, which have the $Q$-scaling at small $\xi$, while the Vainshtein scaling appears as a limiting case of this family of solutions.  
As such, and in contrast to the $Q$-scaling, it never appears as the only possible scaling realized at small $\xi$. 
In the other range  of parameters only singular solution(s) can be found if 
the asymptotic behaviour (\ref{largexibis}) is fixed.

\section{Discussion and conclusions}
\label{CONCLUSIONS}
In this paper we have investigated static spherically symmetric solutions of nonlinear massive gravities in the so-called decoupling limit (DL). After having first identified how to obtain this limit with the ans\"atze we used and which are appropriate for studying static spherical symmetry, we have solved the system of equations left over in the DL and obtained non singular solutions featuring a Vainshtein-like recovery of solutions of General Relativity (GR). This first shows that the singularities found to arise solving the full nonlinear system of equations \cite{Damour:2002gp} are not present in the DL, despite the fact those singularities are commonly thought to be due to a negative energy mode also seen in this limit and associated with the kind of instability first discussed by Boulware and Deser \cite{Boulware:1973my}. To us, this is not necessarily a surprise, since there is in fact no clear clash between having (static), non singular, spherically symmetric solutions and a ghost (coupled to positive energy modes) in a model. In fact, appendix \ref{appG} gives a toy example with such a property. One should of course worry about the possible instabilities (in time) of the static solutions, but this is another story and has a priori nothing to do with the singularities discussed in \cite{Damour:2002gp}.  Moreover, we also found that the scaling at distances smaller than the Vainshtein radius, first conjectured by Vainshtein \cite{Vainshtein:1972sx}, was only a limiting case in an infinite family of non singular solutions each showing a Vainshtein recovery of GR solutions below the Vainshtein radius but a different (but common) scaling at small distances. This new scaling was shown to be associated with zero modes of the nonlinearities left over in the DL. Interestingly, this family of solution all have the same asymptotic power law expansions but differ at small distances which means in particular that the asymptotic expansion is not enough to fix uniquely the solution at small distance. We also noticed, including sources, that the Vainshtein scaling gives a better behaviour at the origin, but it is hard to conclude on this matter without studying the nonlinearities not included in the DL and we plan to come back on this issue in a future work \cite{future}. Last, we have also shown that potentials that were thought not to admit a Vainshtein-like recovery of GR solutions, because the conjectured Vainshtein scaling at small distances would have lead to imaginary solutions, could in fact accommodate such a recovery via a different scaling at small distances. This scaling is the same as the one mentioned above associated with zero modes of the nonlinearities appearing in the DL. 

An intriguing question is whether the properties found in the DL, and summarized above, also hold in the full nonlinear system. This requires in particular solving numerically the full nonlinear system, which is not an easy task given the numerical instabilities we already noticed in the DL. We will carry a thorough analysis of the full nonlinear system, stressing in particular the r\^ole of the neglected interactions, in a future publication \cite{future}. At this stage, let us just say that the findings of this work opens the possibility that important properties of the full nonlinear system could have been overlooked. On the other hand, if indeed this is not the case, it shows that the DL is not capturing all the interesting physics of the spherically symmetric solutions.  It is also important to answer this question for the sake of a better understanding of other models, such as the DGP model where the role of nonlinearities in the equivalent DL and beyond are not fully understood. 
E.g. there are no known non singular exact solution in the bulk and on the brane describing the equivalent of a Schwarzschild Black Hole in DGP theory and results on the Vainshtein mechanism in DGP gravity are all obtained using some approximation scheme \cite{Deffayet:2001uk,APPROX,NONPERT}.

\section*{Acknowledgment}
We thank J.~Garriga, S.~Randjbar-Daemi and especially A.~Vainshtein for interesting discussions. We are also very grateful to A.~Cid, O.~Lopez Pouso and R.~Lopez Pouso for their interest, results and communications with us concerning the differential equation (\ref{wtnew}). We thank much J.~Ecalle for the same and the time taken to communicate with us and explain us his wonderful theory and results.  The work of EB was supported by the EU FP6 Marie Curie Research and Training Network ``UniverseNet'' (MRTN-CT-2006-035863).
 
%\newpage

\begin{appendix}
\section{Equations of motion of $X^A$}
\label{appA}
We first derive the form of the variation $\delta f_{\mu \nu}(x)$ of $f_{\mu \nu}(x)$, defined as in equation (\ref{STUCA}), when $X^A$ varies
as $X^A(x) \rightarrow X^A(x) + \delta X^A(x)$.
We have, replacing $X^A(x)$ by $X^A(x)+\delta X^A(x)$  in the definition of $f_{\mu \nu}(x)$, that $\delta f_{\mu \nu}(x)$ is given by
\ba \label{df}
\delta f_{\mu \nu}(x) &=& \partial_\mu \delta X^A\partial_\nu X^B f_{AB}(X) \nonumber \\
&&+  \partial_\mu X^A \partial_\nu \delta X^B  f_{AB}(X) \nonumber \\
&&+  \partial_\mu X^A \partial_\nu X^B \delta X^C \partial_C f_{AB}(X).
\ea
Using then that
\be
\delta_A^C = \partial_A x^\sigma \partial_\sigma X^C,\nonumber
\ee
where $x^\sigma(X)$ stands for the inverse mapping of $X^A(x)$,
we find that
\be
\partial_\nu X^B f_{AB}(X(x)) = \partial_A x^\sigma f_{\sigma \nu}(x),\nonumber
\ee
and
\be
\partial_C f_{AB}(X) = \partial_C x^\sigma \partial_\sigma X^D \partial_D f_{AB}(X).\nonumber
\ee
Inserting those expression in (\ref{df}), we find that the latter reads
\ba
\delta f_{\mu \nu} &=& \partial_\mu \delta X^A \partial_A x^\sigma f_{\sigma \nu}(x) \nonumber \\
&& + \partial_\nu \delta X^B \partial_B x^\sigma f_{\mu \sigma}(x) \nonumber \\
&&+ \delta X^C\partial_C x^\sigma \partial_\sigma X^D \partial_\mu X^A \partial_\nu X^B \partial_D f_{AB}(X).\nonumber
\ea
We then define $\delta x^\mu(x)$ as in (\ref{defdxmu}), and rewrite $\delta f_{\mu \nu}$ as
\ba \label{STEP1}
\delta f_{\mu \nu} &=& \partial_\mu \delta x^\sigma f_{\sigma \nu} + \partial_\nu \delta x^\sigma f_{\mu \sigma} + \delta x^\sigma \partial_\sigma f_{\mu \nu} \nonumber \\
&& - \delta X^C f_{EF} \partial_\nu X^F\left(\partial_B \partial_Cx^\sigma \partial_\mu X^B \partial_\sigma X^E + \partial_C x^\sigma \partial_\sigma \partial_\mu X^E \right) \nonumber \\
&& - \delta X^C f_{EF} \partial_\mu X^E\left(\partial_A \partial_C x^\sigma \partial_\nu X^A\partial_\sigma X^F +
\partial_C x^\sigma \partial_\sigma \partial_\nu X^F \right)
\ea
If one differentiates with respect to $x^\mu$ (respectively $x^\nu$) the quantity
\be
\delta_C^E = \partial_C x^\sigma \partial_\sigma X^E, \nonumber
\ee
we see that the last terms in the last two lines of the equation (\ref{STEP1}) vanish, and hence
we get that $\delta f_{\mu \nu}$ is given by the expression (\ref{dfmunu}), that is to say the Lie derivative of $f_{\mu \nu}$ along the expression $\delta x^\mu$ considered as a vector field on the space-time manifold. In fact it is easy to see that $\delta x^\mu$ transforms as a vector field under a coordinate change $x'^{\mu} = x'^{\mu}(x)$. Indeed, under such a coordinate change $X^A$ and $\delta X^A$ are scalar quantities, and hence we get that
\be
\delta x'^\mu (x') = \delta x^\sigma (x(x')) \frac{\partial x'^\mu}{\partial x^\sigma},\nonumber
\ee
using the fact that $x'^\mu(X'(x'))$ which appears in the definition of $\delta x'^\mu (x')$ can also be rewritten as
\be
x'^\mu(X'(x')) = x'^\mu(x(X(x(x')))).\nonumber
\ee
We now use the fact that the interaction term between the two metrics, appearing in the action $S_{int}$ is a scalar under reparametrization. Hence, under a coordinate change of the form $x^\mu \rightarrow x^\mu + \xi^\mu$, one has that
\be
 \left(\frac{\delta}{\delta f_{\mu \nu}(x)} {\cal{V}}^{(a)}(f,g)\right)  {\cal L}_{\xi} f_{\mu \nu} +  \left(\frac{\delta}{\delta g_{\mu \nu}(x)}  {\cal{V}}^{(a)}(f,g)\right) {\cal L}_{\xi} g_{\mu \nu} \nonumber
\ee
is a total derivative, so its integral over space-time vanishes (provided we consider asymptotically vanishing $\xi$ fields). Using then that
\be
{\cal L}_{\xi} g_{\mu \nu} = \nabla_\mu \xi_\nu + \nabla_\mu \xi_\nu\,\nonumber
\ee 
and the definition of the energy momentum tensor
(\ref{DEFTMN}), we obtain that under a reparametrization generated by the infinitesimal vector field $\xi^\mu$, one has
\be
\int d^4 x \sqrt{-g} \xi_\mu \nabla_\nu T^{\mu \nu}_g = -\frac{1}{8} m^2 M_P^2 
\int d^4 x  \left(\frac{\delta}{\delta f_{\mu \nu}(x)} {\cal{V}}^{(a)}(f,g)\right) {\cal L}_{\xi} f_{\mu \nu}.\nonumber
\ee
Applying this formula with $\xi^\mu= \delta x^\mu$ defined above, we get the expression (\ref{defin}).

\section{Quadratic part of the equations of motion in the $\lambda, \mu, \nu$ gauge}
\label{appB}
We find for $G_{tt}^{(Q)}$ and $G_{RR}^{(Q)}$ 
\ba
G_{tt}^{(Q)}(\nu,\lambda)&=&-\frac{\lambda^{2}}{2R^{2}}+\frac{\lambda\nu}{R^{2}}-\frac{\lambda\lambda'}{R}+\frac{\nu\lambda'}{R}, \nonumber \\
G_{RR}^{(Q)}(\nu,\lambda)&=&-\frac{\lambda^{2}}{2R^{2}},\nonumber 
\ea 
while the quantities $Q_{tt}$, $Q_{RR}$ and $Q_{b}$ are depending on the choice of interaction term $S_{int}[f,g]$ in action (\ref{action}). 
E.g. for the two interaction terms (\ref{S2}) and (\ref{S3}), they read respectively
\ba
Q_{tt}^{(2)}(\nu,\lambda,\mu)&=&\frac{3\mu^{2}}{2}+\lambda\mu-\frac{9\mu\nu}{4}+R\mu\mu'-\frac{R^{2}\mu'^{2}}{8}-\frac{3\lambda\nu}{4}-\frac{3R\nu\mu'}{4}\nonumber\\
Q_{tt}^{(3)}(\nu,\lambda,\mu)&=&\frac{\lambda^{2}}{4}-\frac{\lambda\nu}{4}+\frac{R\lambda\mu'}{2}+\frac{R^{2}\mu'^{2}}{8}-\frac{3\mu\nu}{4}-\frac{R\nu\mu'}{4}\nonumber\\
Q_{RR}^{(2)}(\nu,\lambda,\mu)&=&\frac{3\lambda\mu}{2}-\frac{3\mu\nu}{4}+\frac{R\mu\mu'}{2}+\frac{3\lambda\nu}{4}+\frac{R\nu\mu'}{4}\nonumber\\
Q_{RR}^{(3)}(\nu,\lambda,\mu)&=&-\frac{3\mu^{2}}{2}+\frac{\lambda\mu}{2}+\frac{\mu\nu}{4}-\frac{R\mu\mu'}{2}+\frac{\lambda\nu}{4}-\frac{R\nu\mu'}{4}-\frac{\nu^{2}}{4}\nonumber\\
Q_{b}^{(2)}(\nu,\lambda,\mu)&=&\frac{\mu'^{2}}{4}+\frac{\mu\mu''}{2}+\frac{2\mu\mu'}{R}+\frac{\nu\mu''}{4}+\frac{\mu\lambda'}{2R}+\frac{\mu'\lambda}{2R}\nonumber\\
&&+\frac{\nu\lambda'}{4R}+\frac{\nu\mu'}{R}-\frac{\mu\nu'}{R}+\frac{\nu\nu'}{R}+\frac{3\lambda\mu}{2R^{2}}+\frac{3\lambda\nu}{2R^{2}}\nonumber\\
\nonumber\\
Q_{b}^{(3)}(\nu,\lambda,\mu)&=&-\frac{\mu'^{2}}{4}-\frac{\mu\mu''}{2}-\frac{2\mu\mu'}{R}-\frac{\nu\mu''}{4}-\frac{\mu\lambda'}{2R}+\frac{\mu'\lambda}{2R}\nonumber\\
&&-\frac{\nu\mu'}{R}-\frac{\nu\nu'}{4R}-\frac{\mu'\nu'}{2}-\frac{\nu\lambda'}{4R}-\frac{\lambda\nu'}{2R}+\frac{\lambda^{2}}{2R^{2}}-\frac{\lambda\nu}{R^{2}}\nonumber
\ea
As we now show, in full generality $Q(\mu)$ depends only on two parameters, $\alpha$ and $\beta$, leading to the expression of Eq.~(\ref{DEFQAB}).
We start from the general expression for the interaction term (\ref{INT})
\be
S_{int}= -\frac{1}{8} m^2 M_{P}^{2} \int d^{4}x \;{\cal V} (g,f)=  -\frac{1}{8} m^2 M_{P}^{2} \int d^{4}x \sqrt{-g} \; V ({\bf g^{-1} f}).\nonumber
\ee
The scalar $V ({\bf g^{-1} f})$ depending only on four invariants made out of the rank two tensor $M^{\mu}_{\nu}\equiv g^{\mu\sigma}f_{\sigma\nu}$ (or using a matrix notation ${\bf M} \equiv {\bf g^{-1} f}$), we choose here those invariants $\Delta_i$ to be given by 
\be
\Delta_{i}\equiv\tr \left[(\bf{M}-\1)^{i}\right],\quad i=1,2,3,4\;.\nonumber
\ee
Thus, one can write
\be
V ({\bf g^{-1} f})=V(\Delta_{i}).\nonumber
\ee
The stress-energy tensor (\ref{DEFTMN}) then reads
\be
T^{g}_{\mu\nu}=-\frac{M_{P}^{2}m^{2}}{4}\left[-\frac{1}{2}\;Vg_{\mu\nu}+\sum_{i=1}^{4}\frac{\delta \Delta_{i}}{\delta g^{\mu\nu}}\;\frac{\partial V}{\partial\Delta_{i}}\right]\nonumber
\ee
where $\frac{\delta \Delta_{i}}{\delta {\bf g}^{-1}}=({\bf M}-\1)^{i}\;{\bf f}$, using again a matrix notation. We can restrict now the metrics to be of the spherically symmetric form (\ref{lammunu}). Since we are interested only in the terms quadratic in  $\mu$, we can further specify our study to the case for which $\lambda=\nu=0$, and expand the stress-energy tensor to the second order in $\mu$, which requires to expand the potential $V$ to third order in $\mu$. This can be done easily noticing that $\Delta_{i}=\mathcal{O}(\mu^{i})$, leading to the expansion:
\be
V(\Delta_{i})=a_{1}+a_{2} \Delta_{1}+a_{3}\Delta_{1}^{2}+a_{4}\Delta_{2}+a_{5}\Delta_{1}^{3}+a_{6}\Delta_{1}\Delta_{2}+a_{7}\Delta_{3}+\mathcal{O}(\mu^{4}).\nonumber
\ee
With this expansion, the Bianchi identity (\ref{BIANCHI}) reads
\ba
\frac{1}{M_{P}^{2}m^{2}}\frac{1}{R}\nabla^\mu T_{\mu R}^g &=&Q^{(\alpha,\beta)}(\mu)-\Omega \left(\frac{\mu'}{2R}+\frac{\mu''}{8}\right) -\frac{3}{80}\;\Omega \;R\mu'\mu''\nonumber \\
&=&0\label{BIANCHIAPP}
\ea
with $Q^{(\alpha,\beta)}(\mu)$ defined as in Eq. (\ref{DEFQAB}), and
\bea
\Omega&=&-a_{2}-4a_{3}-4a_{4}\;,\nonumber\\
\alpha&=&-\frac{1}{24} \Big(a_2  + 4 a_3+16 a_4-48 a_5+24 a_7\Big),\nonumber\\
\beta&=&-\frac{1}{40}  \Big(-7a_2 -68 a_3+88 a_4 -80 a_6-120 a_7\Big).\nonumber
\eea
The last step is to impose the Pauli-Fierz form (\ref{PFMASS}) for the mass term, which fixes the terms linear in $\mu$  in the right hand side of the first line of Eq. (\ref{BIANCHIAPP}) to be zero, i.e. $\Omega=0$. Identifying then the remaining quadratic term $Q^{(\alpha,\beta)}(\mu)$ in (\ref{BIANCHIAPP}) with the quadratic term $Q(\mu_{0})$ in Eq.~(\ref{B2}) leads eventually to the general expression of Eq.~(\ref{DEFQAB}) for the quadratic term in $\mu$ in the Bianchi identity. Hence, as said in the main body of the text, it is here explicitly seen that the fact no term linear in $\mu$ appears in the Bianchi identity, a crucial property for what concerns the scalings appearing in the Vainshtein mechanism, is due to the peculiar tensorial structure of the Pauli-Fierz mass term.

\section{Expansion around $\xi = +\infty$}
\label{LINEARPERT}
Here we study with some details solutions of Eq.  (\ref{Eqy1}) for general $\alpha$ and $\beta$, expanding around the 
leading behaviour (\ref{largexibis}). It is convenient to factorize the asymptotic form of $w$ given by (\ref{largexibis}) and further use the following change of parametrization 
\bea
w(\xi)&=&\frac{2}{3\;\xi^{3}}\left[1-\xi^{3}\;G\left(\Z\right)\right]\nonumber\\
\xi&=&\left(\frac{5\sqrt{2|\beta|}}{3}\Z\right)^{2/5}\; .\nonumber
\eea
The asymptotic behaviour, (\ref{largexibis}) $w\sim w_{\infty}=2/(3\xi^{3})$, translates into
\bea
\xi^{3}\;G(\Z)&\rightarrow& 0\nonumber\\ \label{CondGinf}
\xi^{3}\;\Z G'(\Z)&\rightarrow& 0 
\eea
when $\Z\rightarrow\infty$.
In terms of the new variables $\Z$ and $G$, Eq. (\ref{Eqy1}) takes the form
\ba\
S(\Z)&=&
\left[1-\left(\frac{9\alpha+5\beta}{4\beta}\right)\xi^{3}G(\Z)-\left(\frac{15}{8}\frac{\alpha+\beta}{\beta}\right)\xi^{3}\Z G'(\Z)\right]G''(\Z)\nonumber \\
&&+\left[1-\left(\frac{11}{20}\frac{9\alpha+5\beta}{4\beta}\right)\xi^{3}G(\Z)-\left(\frac{33\alpha+29\beta}{8\beta}\right)\xi^{3}\Z G'(\Z)\right]\frac{G'(\Z)}{\Z}\nonumber \\&&+\frac{|\beta|}{\beta}G(\Z),
\label{EQG}
\ea
with the source $S(\Z)$ being given by 
\be
S(\Z)=\frac{4|\beta|}{\xi^{8}}.\nonumber
\ee
It is then clear that the limiting behaviour (\ref{CondGinf})  for $G$ allows  to drop all the nonlinear terms in $G$ the equation (\ref{EQG}) in the vicinity of infinity. This procedure leads to a Bessel equation of $0$-th order with a source:
\be
G''(\Z)+\frac{G'(\Z)}{\Z}+\frac{|\beta|}{\beta}G(\Z)=S(\Z)\label{BESSEL}.
\ee

For $\beta>0$, the general solution of (\ref{BESSEL}) is given by
\be
G(\Z)=\frac{\pi}{2}J_{0}(\Z)\int_{\Z}^{\infty}Y_{0}(t)S(t)tdt -\frac{\pi}{2}Y_{0}(\Z)\int_{\Z}^{\infty}J_{0}(t)\;S(t)tdt+D_{1}J_{0}(\Z)+D_{2}Y_{0}(\Z)\; .\nonumber
\ee
The only possible choice of integration constants $D_1$ and $D_2$ consistent with the conditions 
(\ref{CondGinf}), is  that they both vanish, i.e. $(D_1,D_2)=(0,0)$. In this case, $G$ has the asymptotic behaviour $G(\Z)=\mathcal{O}(\Z^{-16/5})$ and the asymptotic conditions (\ref{CondGinf}) fixes the solution uniquely. 
This is similar to the example (\ref{example1}) discussed in the text and also allows to understand qualitatively properties of the numerical integration of equation (\ref{Eqy1}) for positive $\beta$. Indeed, in this case, if one integrate numerically inwards (from large $\xi$), numerical errors can source the homogeneous modes $J_{0}(\Z), Y_{0}(\Z)$. However, the later are not growing out of control in agreement with the fact the numerical solution is found to be stable under such small perturbations.

For $\beta<0$, the general solution of (\ref{BESSEL}) reads
\be
G(\Z)=-K_{0}(\Z)\int_{1}^{\Z}I_{0}(t)S(t)tdt -I_{0}(\Z)\int_{\Z}^{\infty}K_{0}(t)\;S(t)tdt+D_{3}K_{0}(\Z)+D_{4}I_{0}(\Z)\; .\nonumber
\ee
It this case, the conditions (\ref{CondGinf}) only leads to the vanishing of $D_4$, but the other integration constant $D_3$ remains free, since the homogeneous mode $K_{0}(\Z)$ decays fast enough at infinity. Whatever the chosen $D_3$, one finds a leading behaviour for $G$ given by $G(\Z)=\mathcal{O}(\Z^{-16/5})$. This is analogous to the example (\ref{example2}) discussed in the main body of the text. The existence of the growing mode $I_{0}(\Z)$ allows us to understand better the numerical instabilities observed in the case of the \AGS\ potential. It is indeed impossible not to source this growing mode numerically, and, after some point, this mode dominates the solution and the integration reaches a singularity. As a consequence, one has to tune very carefully the initial conditions of the numerical integration in order to avoid this explosion.

\section{An example of a system with coupled normal and ghost fields.}

Let us consider the following higher-order derivative action, $S$, for the scalar field $\phi$
\be
S=\int d^4 x \left\{\frac12\left(\Box\phi\right)^2+T\phi\right\}.
\label{actionB2}
\ee
The above system can be described in terms of two fields, one positive energy ("normal") field, $\phi_c$, and one negative energy (ghost) field, $\psi_c$. In terms of these new variable, $S$ now reads 
\be
S=\int d^4 x \left\{\frac12\phi_c\Box\phi_c-\frac12\psi_c\Box\psi_c-\frac14\left(\phi_c-\psi_c\right)^2
+\frac{T}{\sqrt2}\left(\phi_c+\psi_c\right)\right\},
\ee
where it is explicitly seen that the normal and ghost degrees of freedom are 
coupled via a potential term.
From (\ref{actionB2}) we find the equation of motion for $\phi$,
\be
\Box\left(\Box\phi\right)=T.
\label{eomB2}
\ee
In the case of the point-like source, the full solution of (\ref{eomB2})
is,
\be
 \phi(R)= \frac{A_1 R}{2}+\frac{A_0 R^2}{2}-\frac{B_1}{R}+B_0.
\label{solB2}
\ee
The first term in (\ref{solB2}) ``kills'' the delta-function in the r.h.s. of (\ref{eomB2}).
Note, that the solution is smooth everywhere for $R>0$. 
Moreover, the presence of the growing with $R$ terms in (\ref{solB2}) is
not dangerous, since these terms do not lead to a ``bad'' physical behaviour
at $R\to \infty$. Indeed, the energy density, $E(R)$, for the configuration of $\phi$ given 
by (\ref{solB2}) is \cite{Anisimov:2005ne}
\be
 E(R)=\frac12\left(A_0+\frac{A_1}{R}\right)^2
	+\frac{A_1}{R^2}\left(\frac{A_1}{2}+A_0R+\frac{B_1}{R^2}\right).
\label{energyB2}
\ee
Thus, $E(R)\to (1/2)A_0^2$ as $R\to \infty$.
\label{appG}

\end{appendix}

\end{document}